\documentclass{aa}  

\usepackage{graphicx}
\usepackage{txfonts}
\usepackage{hyperref}
\usepackage[table]{xcolor}
\newcolumntype{C}[1]{>{\centering\arraybackslash}m{#1}}
\usepackage{array}

\begin{document} 

    \title{That's no moon: Exomoon survival and detection limitations}

    \author{Alicia Pérez-Rodrigo\inst{1}\email{aliper09@ucm.es}
          \and Isabel Rebollido \inst{2,3} \corrauth{irebollido@cab.inta-csic.es}
    }
    \institute{
Departamento de Astrof\'isica, Universidad Complutense de Madrid, 28040 Madrid, Spain
\and
European Space Agency (ESA), European Space Astronomy Centre (ESAC), Camino Bajo del Castillo s/n, 28692 Villanueva de la Cañada, Madrid, Spain
\and
Centro de Astrobiología (CSIC-INTA), Camino Bajo del Castillo s/n, 28692 Villanueva de la Cañada, Madrid, Spain
}

    \date{Received June 04, 2026; accepted August 15, 2026}

\abstract
    {While moons are ubiquitous in the Solar System, exomoons have not
yet been unambiguously confirmed despite the discovery of over 5,600 exoplanets.}
    {We investigated the reasons for this lack of detections by
reviewing high-impact literature searches, assessing dynamical stability and detectability constraints, and comparing these with the moon population in the Solar System.}
    {We compiled a homogeneous catalogue of 24 proposed exomoon
candidates, including stellar, planetary, and inferred lunar parameters. For each case we applied Hill, Roche, and Laplace stability criteria and estimated detectability limits for current and forthcoming instruments. In addition, we tested 6038 confirmed exoplanets against the same stability and mission-specific observational filters for a given range of moon parameters.}
     {A large fraction of the potential exomoon candidates investigated in the literature are either inconsistent with long-term stability or produce signals below Kepler sensitivity. Only 8 of the 13 candidates meet at least one of the stability criteria. Our extended analysis highlights 27 promising targets for exomoon searches with PLATO, while moons with Solar-System-like properties remain below current thresholds.}
    {The lack of exomoon detections is likely due to a combination of the inefficient formation of stable massive moons and limited instrumental sensitivity. Other processes, such as early moon loss during planetary migration or giant impacts, may also influence moon survival. Future searches should prioritise systems that combine long-term dynamical stability with favourable observational conditions for next-generation photometric and astrometric facilities.}


    \maketitle
    \nolinenumbers

\section{Introduction}
\label{intro}

Natural satellites are common throughout the Solar System, with nearly 300 moons identified to date. Only Mercury and Venus lack moons, likely due to their proximity to the Sun and dynamical constraints \citep{Sheppard_2009}. Moons provide valuable records of the dynamical and formation history of their host planets \citep{joy_2016_the, barr_2016_formation, heller_2014_formation}, and their potential role in stabilising planetary obliquity and supporting habitable conditions has further motivated interest in detecting such bodies beyond the Solar System \citep{lammer_2014_origin}. 

Several formation pathways have been identified from Solar System examples. Co-formation within circumplanetary disks typically produce prograde, low-inclination satellites with characteristic moon-to-planet mass ratios of $M_{\rm m}/M_p \sim 10^{-4}$, a value remarkably consistent across the Jovian, Saturnian, and Uranian systems 
\citep{heller_2014_formation,Shibaike2025}. Capture processes, often via binary exchange, explain irregular or retrograde orbits such as 
those of Triton and possibly Mars’s moons \citep{refId0}. Giant impacts can yield unusually large satellites, with the Earth--Moon system providing the canonical example where $M_m/M_p \gtrsim 10^{-2}$.

More speculative mechanisms include streaming instability in massive, gas-poor disks, which may generate small moonlets \citep{nakajima_2024_the}. Thermal constraints from the ice-rich composition of Ganymede and Callisto further suggest that icy moons can only form once circumplanetary accretion rates decline to $\dot M_g \lesssim 0.2\,M_J\,\mathrm{Myr}^{-1}$, where $M_J$ denotes one Jupiter mass. Recent simulations of satellite formation in impact-generated disks \citep{Madeira2025} show that the mass of the largest moon scales nearly linearly with the initial disk mass. A threshold of $M_{\rm disk} \sim 0.03\,M_p$ is required to form a single dominant satellite, whereas lower-mass disks typically yield multiple small moons. These results suggest that massive Earth--Moon analogues are rare, implying that most exomoon systems may remain below current detection limits.

Yet, formation alone does not guarantee survival. The dynamical stability of moons is constrained by perturbations from the host star and planets in the system (usually quantified by the Hill radius and Roche limit), which together restrict the parameter range of long-term stable orbits. Moons orbiting too close to their host planets risk tidal disruption, whereas those too far away may be stripped by stellar perturbations. These stability considerations, developed both from analytic arguments and numerical simulations, are critical for interpreting the survival rate of exomoons in different planetary architectures \citep{barnes_2002_stability, hamers_2018_stability, rosariofranco_2020_orbital, dobos_2021_survival, makarov_2023_pathways}.

Despite the discovery of over 5,600 exoplanets, no exomoon has yet been confirmed \citep{kipping_2009_on, teachey_2018_evidence,Teachey_2020}. Several observational techniques have been proposed to search for such companions, including transit timing variations (TTVs) \citep{kipping_2009_transit}, direct imaging \citep{kenworthy_2015_modeling} and microlensing \citep{miyazaki_2018_moa2015blg337}. Dedicated searches, such as the Hunt for Exomoons with Kepler (HEK) project \citep{Kipping:2012vt}, have yielded controversial candidates (e.g. Kepler-1625b-i) that remain unconfirmed \citep{teachey_2018_evidence, kipping_2024_a, heller_2024_large}.
Ongoing developments, including machine-learning approaches \citep{weghs_2021_photometric, teachey_2021_identifying} and the capabilities of future missions such as Habitable Worlds Observatory (HWO) \citep{limbach_2024_exomoons} or PLAnetary Transits and Oscillations of stars (PLATO)\citep{plato}, promise to improve the prospects for exomoon detection in the coming decades.

This work aims to place current knowledge of exomoon formation, dynamics, and detection into a unified context to better understand the challenges associated with their non-detections. Section~\ref{sec: catalogue} introduces our catalogue of proposed candidates and analogous exomoon systems reported in the literature. Section~\ref{sec:stability_criteria} presents the dynamical stability criteria applied to these systems. Section~\ref{sec: observational_limits} explores observational constraints and the sensitivities of current and future missions. Section~\ref{sec:disruption} discusses additional disruption mechanisms affecting long-term moon survival. Section~\ref{sec:comparative} compares the candidate population with moons in the Solar System. Finally, Sect.~\ref{sec:conclusion} summarises our main results and future prospects.

\section{Data and catalogue construction}
\label{sec: catalogue}
We compiled a list of 24 proposed exomoon candidates from the literature published between 2009 and 2024, based on observational features such as timing anomalies, light curve residuals, and dynamical perturbations that may indicate the presence of an exomoon. The compilation includes systems with published observational support as well as cases reported in the literature where such signatures have been suggested, even if they are not formally classified as exomoon candidates. For simplicity, throughout this work we refer to all of these systems as exomoon candidates, regardless of their current status.

The naming convention commonly used in the literature involves appending a Roman numeral (e.g. `i' for the first moon and `ii' for the second) to the name of the host planet. We adopt this convention throughout the paper. In addition, values reported without uncertainties or confidence intervals correspond to estimates derived from observational data or theoretical models.

The catalogue is hosted online in a public \href{https://github.com/Aliper02/Exomoons_Catalogue/}{GitHub repository} that will remain open to updates as new observations emerge, particularly from current and upcoming missions such as James Webb Space Telescope (JWST) \citep{gardner_2006_the}, CHaracterising ExOPlanet Satellite (CHEOPS) \citep{2021_CHEOPS}, or PLATO \citep{PLATO_2017}. An extract with the main physical and orbital parameters is presented in Table~\ref{tab:full_catalogue}.

\begin{table*}
\caption{Catalogue of 24 putative exomoon candidates and their characteristics.}
\label{tab:full_catalogue}
\centering
\renewcommand{\arraystretch}{1.0}
\scriptsize
\resizebox{\textwidth}{!}{
\begin{tabular}{l l c l c l c c}
\hline\hline
Exomoon & Detection method & $M_m$ ($M_J$) & Host star & $M_\star$ ($M_\odot$) & Host planet & $M_p$ ($M_J$) & Ref. \\
\hline
Kepler-1625 b-i         & Primary transit, 2017          & 0.06                      & Kepler-1625                & 0.96  & Kepler-1625 b         & 11.6     & 1 \\
MOA-2011-BLG-262L b-i   & Microlensing, 2014             & 0.00148                   & MOA-2011-BLG-262L          & 0.19  & MOA-2011-BLG-262L b   & 0.09099  & 2 \\
2MASS J1119-1137 b-i    & Primary transit, 2021          & --                        & System without star        & --    & 2MASS J1119-1137 b    & 3.7      & 3 \\
Kepler-1708 b-i         & Primary transit, 2021          & 0.116                     & Kepler-1708                & 1.088 & Kepler-1708 b         & 4.6   &   1 \\
Kepler-1513 b-i         & TTV, 2022                      & 0.0082                    & Kepler-1513 A              & 0.943 & Kepler-1513 Ab        & 0.152    & 4 \\
Beta Pic b-i      & Obliquity analysis             & 0.053                     & Beta Pic                   & 1.73  & Beta Pic b            & 11.90    & 5 \\
J1407b-i                & Ring system analysis           & $<$0.0025                 & 1SWASP J1407               & 0.9   & 1SWASP J1407b         & 20       & 6 \\
2MASS J2117-2940-i      & Primary transit, 2024          & 0.0016                    & No star                    & --    & 2MASS J2117-2940      & 7        & 7 \\
DH Tau-i                & Direct imaging, 2005           & 1.0002                    & DH Tau                     & 0.37  & DH Tau b              & 11       & 8 \\
HD 189733 Ab-i          & Primary transit, 2014          & --                        & HD 189733 A                & 0.8   & HD 189733 Ab          & 1.138    & 9 \\
HD 189733 Ab-ii         & Other, 2019                    & 0.00005                   & HD 189733 A                & 0.8   & HD 189733 Ab          & 1.138    & 10 \\
Kepler-409b-i           & TTV, 2020                      & 0.0009                    & Kepler-409                 & 0.92  & Kepler-409b           & 0.0692   & 11 \\
Kepler-517Ab-i          & Primary transit, 2020          & 0.0016                    & Kepler-517 A               & 0.991 & Kepler-517 Ab         & 0.024    & 11 \\
Kepler-809b-i           & Primary transit, 2020          & 0.0092                    & Kepler-809                 & 0.99  & Kepler-809b           & 0.12     & 11 \\
Kepler-857b-i           & Primary transit, 2020          & 0.0051                    & Kepler-857                 & 0.99  & Kepler-857b           & 0.044    & 11 \\
Kepler-1000b-i          & Primary transit, 2020          & 0.0049                    & Kepler-1000                & 1.40  & Kepler-1000b          & 0.063    & 11 \\
Kepler-1326b-i          & Primary transit, 2020          & 0.0191                    & Kepler-1326                & 1.45  & Kepler-1326b          & 0.077    & 11 \\
Kepler-1442b-i          & Primary transit, 2020          & 0.005                     & Kepler-1442                & 1.34  & Kepler-1442b          & 0.044    & 11 \\
KOI-268.01-i            & Primary transit, 2020          & 0.0026                    & KOI-268                    & 1.18  & KOI-268.01            & 0.029    & 11 \\
MOA-2015-BLG-337a-i     & Microlensing, 2018             & 0.106                     & No star                    & --    & MOA-2015-BLG-337 a    & 9.8      & 12 \\
WASP-12b-i              & Primary transit, 2014          & --                        & WASP-12 A                  & 1.434 & WASP-12 Ab            & 1.465    & 12 \\
WASP-49b-i              & Other, 2019                    & 0.00005                   & WASP-49 A                  & 0.989 & WASP-49 Ab            & 0.399    & 12 \\
WASP-76 Ab-i            & Other, 2019                    & 0.00005                   & WASP-76 A                  & 1.46  & WASP-76b              & 0.920    & 13 \\
WASP-121b-i             & Other, 2019                    & 0.00005                   & WASP-121                   & 1.353 & WASP-121b             & 1.184    & 13 \\
\hline
\end{tabular}
}
\tablefoot{This table lists the names of exomoon candidates along with their proposed detection method, estimated exomoon mass ($M_m$), and basic parameters of the host star and planet. Stellar masses are given in solar masses ($M_\odot$) and planetary and exomoon masses in Jupiter masses ($M_J$). References are provided in the last column. The complete table with all the parameters is available in the \href{https://github.com/Aliper02/Exomoons_Catalogue/blob/main/Candidate_exomoons_github.xlsx}{GitHub repository}. For entries (1)--(5), the most recent paper available on each system is cited, although the numerical and physical parameters were retrieved from the \href{https://exoplanet.eu}{exoplanet.eu} database.}

\tablebib{
(1)~\citet{kipping_2024_a};
(2)~\citet{Bennett_2014};
(3)~\citet{limbach_2021_on};
(4)~\citet{Michel_2024};
(5)~\citet{Poon_2024};
(6)~\citet{kenworthy_2015_modeling};
(7)~\citet{limbach_2024_occurrence};
(8)~\citet{lazzoni_2020_the};
(9)~\citet{benjaffel_2014_transit};
(10)~\citet{oza_2019_sodium};
(11)~\citet{fox_2020_exomoon};
(12)~\citet{miyazaki_2018_moa2015blg337};
(13)~\citet{seidel_2019_hot}.
}
\end{table*}

\subsection{Sources and compilation methodology}

According to \citet{kipping_2020_an}, identifying an exomoon candidate requires the presence of significant TTVs, a detectable periodicity in those variations, and a derived moon mass significantly different from zero. Not all systems studied in the literature satisfy these criteria; therefore, only a subset is considered as strong exomoon candidates. A key reference for this compilation is the sample presented in \citet{kipping_2020_an}, which includes six planetary systems with TTVs analysed under the hypothesis of an exomoon.

All available parameters are provided in the \href{https://github.com/Aliper02/Exomoons_Catalogue/blob/main/Candidate_exomoons_github.xlsx}{GitHub repository}: exomoon status, detection method, moon mass ($M_m$) and moon semi-major axis ($a_m$), host star characteristics (name, mass ($M_\star$), and effective temperature, $T_{\text{eff}}$), and planetary properties (name, mass ($M_p$), semi-major axis ($a_p$), and radius ($R_p$)). Exomoons are classified internally as candidates, tentative detections (not yet confirmed), or controversial cases where the interpretation is not universally accepted (e.g. \citet{kipping_2024_a}).

Exomoon candidates have been proposed using a variety of indirect detection techniques. The most common approach relies on TTVs and transit duration variations (TDVs), where periodic deviations in the planet's transit are interpreted as the gravitational influence of a companion satellite \citep{kipping_2009_transit, kipping_2009_transit1, kipping_2013_the}. However, alternative approaches have also been used to explain possible candidates. For instance, extended ring structures may indicate the presence of exomoons, as in J1407b \citep{kenworthy_2015_modeling} (see Subsection~\ref{special}); axial tilt measurements and their misalignment have been used to infer exomoon presence in systems such as Beta Pictoris b \citep{Poon_2024}; and indirect evidence of volcanic satellites has been suggested through transmission spectroscopy of alkali metals, such as sodium or potassium, in exoplanetary atmospheres, as proposed for HD~189733 Ab-i \citep{benjaffel_2014_transit}, WASP-49b-i \citep{miyazaki_2018_moa2015blg337}, WASP-76 Ab-i \citep{seidel_2019_hot}, and WASP-121b-i \citep{seidel_2019_hot}.

A few systems discussed in \cite{teachey_2024_detecting} were not included in our main catalogue for specific reasons. We excluded cases where circumplanetary material has been detected but no exomoon candidate has been inferred. For instance, PDS 70 c hosts a confirmed circumplanetary disk, which might represent an active site of moon formation but lacks any observational signature that could be attributed to an exomoon. Similarly, OGLE-2015-BLG-1459L, exhibits degenerate microlensing solutions where the exomoon interpretation is disfavoured. 2MASS J1147--2040, studied alongside with 2MASS J1119--1137, showed no definitive dimming event attributable to a satellite. Lastly, K2-33 b has been suggested to host circumplanetary rings rather than moons. These cases are therefore excluded from the catalogue, although they remain of interest for future observations.

Current exomoon searches are strongly biased towards large satellites orbiting massive planets and relatively quiet host stars, particularly in systems with long observational baselines and high signal-to-noise (S/N) photometry. Consequently, the presently proposed exomoon candidates are unlikely to represent the true underlying exomoon population, but rather the subset detectable with current observational techniques.

\subsection{Special systems: Brown dwarfs and ring systems}
\label{special}
Several exomoon candidates in Table \ref{tab:full_catalogue} orbit brown dwarfs or objects suspected to be brown dwarfs. While brown dwarfs can themselves be companions to stars, they may also host planetary-mass companions in circumsubstellar orbits. However, the terminology is not always consistent in the literature, and classification may vary depending on the system's formation history \citep{heller_2014_formation}. In this work, four systems included in our \href{https://github.com/Aliper02/Exomoons_Catalogue/blob/main/Candidate_exomoons_github.xlsx}{GitHub} catalogue involve exomoon candidates orbiting brown dwarfs with large mass uncertainties: 2MASS J1119–1137 b, 2MASS J2117–2940, DH Tau b, and J1407b. 

When looking at exomoon estimations, the objects WASP-49b-i, WASP-76 Ab-i, and WASP-121b-i exhibit low masses, of the order of $10^{-5}~M_J$. While still over three orders of magnitude more massive than Saturn’s rings ($\sim10^{-8}~M_J$) \citep{iess_2019_measurement}, such values could be consistent with massive or dense ring structures rather than true satellites. This is particularly plausible in systems hosting young or giant exoplanets, where circumplanetary rings may be more substantial. Indeed, WASP-76 Ab and WASP-121b are inflated hot Jupiters, while WASP-49b is more accurately classified as a hot Saturn, although they are not particularly young (approximately gigayear-aged systems). In contrast, the proposed ring system around the substellar companion J1407b, a much younger object ($\sim$16 Myr), is estimated to have a total mass between $0.01$ and $0.1~M_J$, i.e. $10^{6}-10^{7}$ times more massive than Saturn’s rings. Its complex light curve, interpreted as the transit of a colossal ring system, includes a gap that may indicate the presence of a forming moon. This proposed satellite, J1407b-i, is therefore included in our catalogue as an illustrative case linking massive ring structures to potential satellite formation.

\section{Stability criteria for exomoon candidates}
\label{sec:stability_criteria}

The absence of confirmed exomoon detections may be related to dynamical processes affecting their long-term stability. Strong stellar tidal forces can significantly perturb moons orbiting planets with small semi-major axes. Moreover, simulations have shown that planetary sizes influence these outcomes \citep{makarov_2023_pathways}. In addition, the relative rotation between the planet and moon can drive inward or outward migration through tidal interactions \citep{heller_2014_formation}. In this section, we explore several criteria that constrain the dynamical stability of the candidate systems.

\subsection{Hill radius}
\label{subsec: Hill}
The Hill Radius, $r_H$, defines the region around a planet where its gravitational influence dominates over that of the host star, allowing it to retain satellites. For an exomoon to remain gravitationally bound and dynamically stable, its apocentre must lie within a more restrictive value known as the reduced Hill Radius, $r_H'$. The stability condition incorporates long-term stability considerations, such as orbital eccentricity and direction of motion, and is essential for assessing whether a moon can survive tidal and dynamical interactions in a three-body system:

\begin{equation}
    a_m (1 + e_m) < r'_H = B a_p (1 - b e_p) \left( \frac{M_p}{3 M_*} \right)^{1/3},
\end{equation}

where $e_p$ is the eccentricity of the planet, $e_m$ is the eccentricity of the moon, and the constants $B$ and $b$ depend on the moon's orbital configuration. For prograde orbits (same direction as the planet’s orbit): $B = 0.49$, $b = 1.03$, and for retrograde orbits (opposite direction): $B = 0.93$, $b = 1.08$. 
The expression above follows the empirical prescription of \citet{Domingos2006}, which has become the standard analytical approximation for long-term exomoon stability and has been adopted in several recent studies (e.g. \citealt{makarov_2023_pathways}). More recent long-term $N$-body simulations indicate that the classical Hill approximation may introduce non-negligible deviations for systems with unusually large planet-to-star mass ratios, where the exact three-body equilibrium points depart from the simplified Hill solution \citep{barbosa_2026_on}. Nevertheless, for the planetary systems considered in this work, the Domingos et al. prescription provides a suitable first-order analytical estimate of the stability boundary.

Compared to the classical Hill radius, which only considers the planet-star mass ratio, the reduced Hill radius provides a more realistic and conservative estimate of the region where moons can remain stable over long timescales. It accounts for tidal interactions and dynamical instabilities shown in simulations. In fact, studies such as \citet{makarov_2023_pathways} suggest that only a small fraction (about 10--26$\%$) of known exoplanets can host stable moons, primarily those with sufficiently large reduced Hill spheres.

In general, prograde moons are stable within approximately 40\% of the Hill radius. In contrast, retrograde moons can remain stable at greater distances due to different tidal dynamics and weaker resonant perturbations \citep{rosariofranco_2020_orbital}. 
The implications of these limitations and their comparison with recent numerical stability studies are further discussed in Sect.~\ref{sec:disruption}.

\subsection{Roche radius}
\label{subsec: Roche}
The Roche Radius, $r_R$ \citep{Roche1850,Chandrasekhar1963}, defines the minimum distance at which a moon can orbit a planet without being destroyed by tidal forces. It sets a fundamental constraint on the inner edge of a stable exomoon orbit. To avoid tidal disruption, the pericentre of the moon’s orbit, the closest point to the planet, must remain beyond the Roche radius.

This condition is expressed as
\begin{equation}     
    a_m(1 - e_m) > r_R = 2.2 R_p \left( \frac{M_p}{M_m} \right)^{1/3}.
\end{equation}
If this condition is not met, the tidal forces exerted by the planet exceed the moon’s self-gravity, leading to structural disintegration. Together, the Roche limit and the reduced Hill radius define the classical inner and outer boundaries within which a satellite can remain both gravitationally bound and tidally intact. However, these criteria do not explicitly account for the long-term secular perturbations induced by the host star, which may affect orbital stability even for satellites formally located within the Hill sphere.
 
\subsection{Laplace criterion}
\label{subsec: Laplace}
Although the reduced Hill criterion provides the primary constraint on the outer extent of stable satellite orbits, we also consider the Laplace criterion as a complementary diagnostic. Unlike the Hill limit, which describes the boundary for gravitational retention, the Laplace criterion characterises the regime where stellar perturbations become dynamically important and may drive long-term secular evolution of the satellite orbit. We therefore used it as an additional filter rather than as an alternative stability criterion.

Stellar perturbations can induce secular resonances, which are long-term gravitational interactions that gradually modify the moon’s orbital elements. The larger the semi-major axis of the moon's orbit, the greater the influence of the star's gravitational pull.

Secular resonances occur when the precession rate of the moon's orbit matches that of another body in the system. Over long timescales (typically 10-10$^3$~Myrs), these resonances can increase the moon’s eccentricity or inclination, potentially leading to orbital instability, collision, or ejection \citep{malhotra_1998_orbital, mary_solar}.

An empirical approximation for the outer boundary of orbital stability, known as the Laplace limit, is given by
\begin{equation}     
a_m < 0.5\, a_p \left( \frac{M_{\star}}{M_p} \right)^{1/3}. 
\end{equation}
This criterion provides a practical upper limit for stable moon orbits by accounting for long-term stellar perturbations and their role in driving secular evolution.

\subsection{Tidal evolution}

Beyond purely geometric constraints (Hill, Roche, and Laplace), tidal evolution further restricts the long-term survival of planet--moon systems. A recent analytic study \citep{Makarov2025} derived a universal condition for tidal synchronisation: 
\begin{equation}
C = \Omega_p(t_0) + X\sqrt{a(t_0)} \;\; \geq \;\; C_{\rm syn},
\end{equation}
where $\Omega_p(t_0)$ is the initial spin rate of the planet, $a(t_0)$ the initial moon semi-major axis, and $X$ a constant that depends on the planetary and lunar masses and the planetary radius. If this inequality is not satisfied, the moon is ultimately lost through orbital decay or escape. When fulfilled, two synchronous states exist: an inner unstable configuration and an outer stable equilibrium. These results, confirmed across several Solar System examples, suggest that massive long-lived exomoons are expected only in a narrow region of parameter space, reinforcing their intrinsic rarity.

Numerical simulations indicate that close-in planets with orbital periods shorter than 10 days are unlikely to retain stable moons due to strong tidal perturbations from the host star \citep{dobos_2021_survival}. For planets with wider orbits (10–300 days), exomoon survival rates improve significantly, reaching values between 70--90$\%$ depending on system architecture.

Although tidal evolution plays a fundamental role in long-term satellite survival, applying detailed tidal models requires parameters that are currently unavailable for most catalogued systems. These parameters include tidal quality factors and internal dissipation properties.

\subsection{Stability maps}

We can apply the former criteria to the exomoon candidates in Table \ref{tab:full_catalogue}. We will use them to investigate whether the presence of natural satellites is compatible with the proposed system architectures.

\begin{figure*}
    \centering
    \includegraphics[width=\hsize]{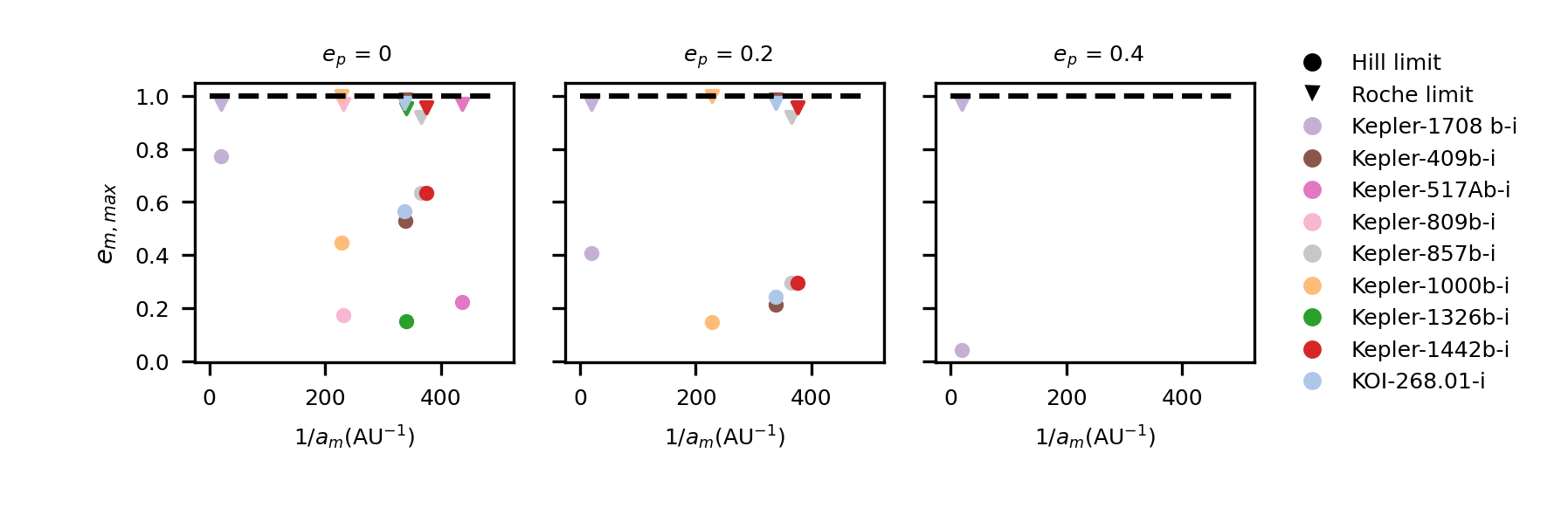} 
    \caption{Stability maps for the subset of exomoon candidates with well-constrained parameters (i.e. systems with available planetary and putative moon masses, as well as semi-major axes). The figure illustrates the maximum allowed eccentricity of the moon ($e_{m,\mathrm{max}}$) as a function of the inverse of the semi-major axis ($1/a_m$). The three panels correspond to different assumed eccentricities of the host planet ($e_p = 0$, $0.2$, and $0.4$ from left to right). The horizontal dashed line at $e_{m,\mathrm{max}}=1$ marks the maximum physically allowed eccentricity. Stability thresholds are defined by two criteria: the Hill criterion ($e_m < r_\mathrm{H}'/a_m - 1$), represented by circular dots, and the Roche criterion ($e_m < 1 - r_\mathrm{R}'/a_m$), represented by inverted triangles. Each coloured point corresponds to an individual exomoon candidate. For $e_p = 0$, eight candidates can be stable; for $e_p = 0.2$, six remain stable; and for $e_p = 0.4$, only one candidate is stable. This representation highlights how increasing planetary eccentricity strongly reduces the available stable region.}
    \label{fig:stability criteria}
\end{figure*}

Given that there are no estimations for exomoon eccentricities and that not all discovered exoplanets have reliable eccentricity values, we calculated the upper limit of the moon eccentricity, $e_m$, for three representative values of the planetary eccentricity, $e_p$, using both the Hill and Roche criteria. This allowed us to study the stability of exomoons under various orbital conditions and characterise the situations where the moon might remain in a stable orbit over time. The first results of these calculations, applied only to the candidate exomoons for which sufficient orbital data were available, are presented in Fig.~\ref{fig:stability criteria}. 

For \(e_p = 0\), we find that 5 out of 13 moons in the catalogue are not stable under either the Hill or Roche condition. For the case of $e_p=0.4$, only one moon could be stable at very low values of eccentricity, while for $e_p=0.5$, no moons would be stable at all. These criteria are static, so they do not account for orbital evolution or perturbations over time. Configurations that violate these limits are expected to become dynamically unstable on mid-to-long timescales ($10^3$–$10^6$ yr), although they may appear temporarily stable in the short term. Precise timescales for orbital disruption or ejection can only be determined through numerical simulations.

\begin{figure}
    \centering
    \includegraphics[width=10.9cm]{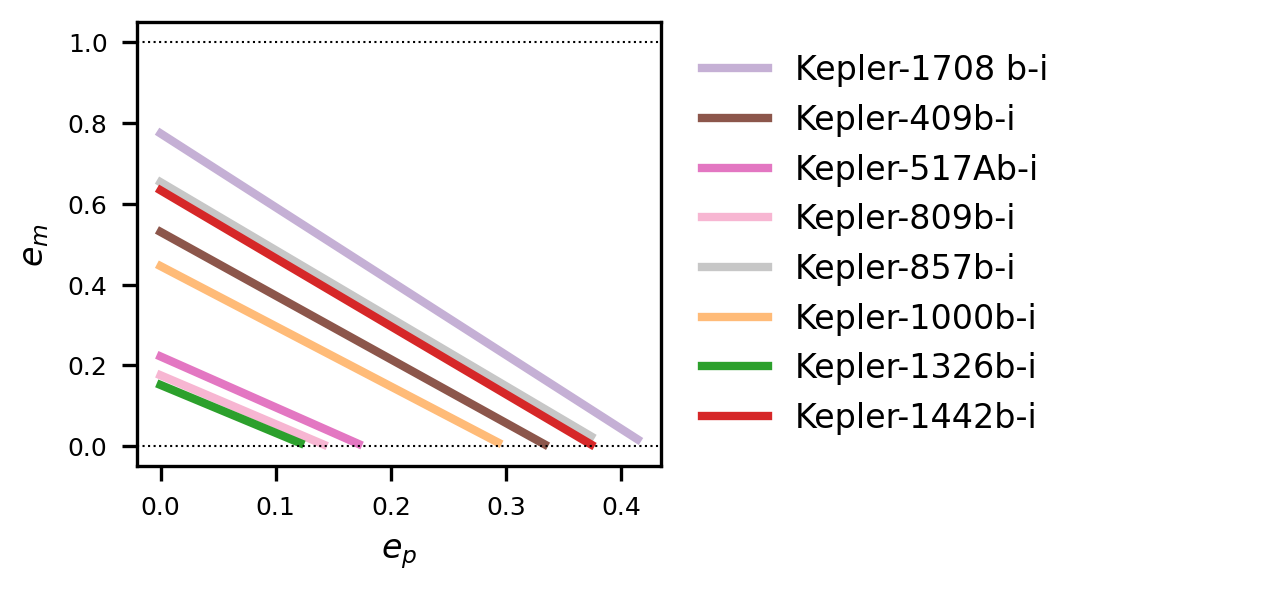}
     \caption{Maximum allowed moon eccentricity ($e_m$) as a function of the planetary eccentricity ($e_p$), derived from the Hill stability criterion. 
    Each coloured line corresponds to a different exomoon candidate, with the planetary and lunar parameters ($M_p$, $a_p$, $a_m$) fixed for that system. 
    The curves were obtained by varying $e_p$ and computing the maximum $e_m$ that satisfies the Hill stability condition, 
    $e_m < r_H'/a_m - 1 = B (a_p / a_m) (1 - b e_p) (M_p / 3 M_*)^{1/3} - 1$. 
    The upper and lower horizontal dotted lines indicate the physical limits $e_m = 1$ and $e_m = 0$, respectively. 
    The differences between the lines reflect the diversity in planet masses and orbital configurations in the catalogue. 
    The figure shows that the maximum stable eccentricity of a moon decreases rapidly with increasing planetary eccentricity, 
    with stability windows closing for $e_p \gtrsim 0.3$--$0.4$, highlighting the strong dynamical constraint imposed by planetary eccentricity on exomoon survival.}
    \label{fig:em_vs_ep}
\end{figure}

In Fig.~\ref{fig:em_vs_ep}, we show the maximum allowed eccentricity of the moon ($e_m$) as a function of the planetary eccentricity ($e_p$), derived from the Hill stability criterion. To compute these curves, we applied the condition

\begin{equation}
    e_m < \frac{r_H'}{a_m} - 1 = B \frac{a_p}{a_m} (1 - b e_p) \left( \frac{M_p}{3 M_*} \right)^{1/3} - 1,
\end{equation}

using the known semi-major axes ($a_p$ and $a_m$) and masses ($M_p$ and $M_*$) for each planet--moon system in our catalogue. Each coloured line corresponds to a different exomoon candidate, with the planetary and lunar parameters fixed for that system. By varying $e_p$ from 0 to 0.5, we derived the maximum allowed moon eccentricity $e_m$ that ensures Hill stability. 

The slope of each curve in Fig.~\ref{fig:em_vs_ep} depends on the quantity 
$A = B(a_p/a_m)(M_p/3M_*)^{1/3}$, such that
$d e_m / d e_p = -Ab$. 
As a result, systems with larger $a_p/a_m$ ratios or more massive planets exhibit steeper declines in the maximum stable moon eccentricity as $e_p$ increases, while systems with smaller ratios retain wider stability regions.

\begin{figure}
    \centering
    \includegraphics[width=\hsize]{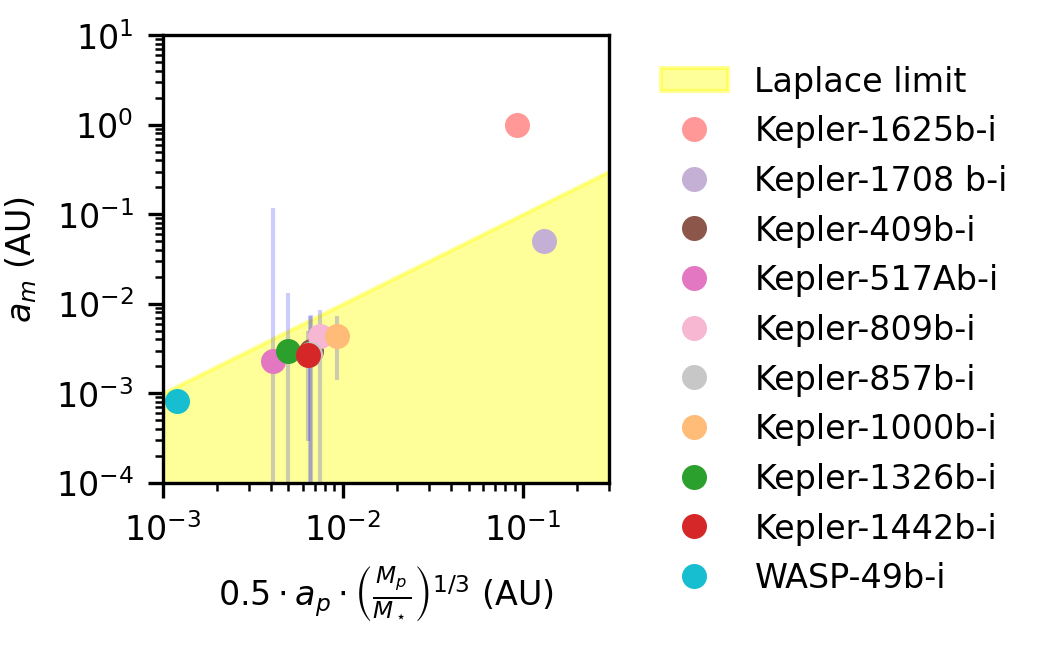} 
   \caption{Test of the Laplace stability criterion for exomoon candidates. 
    The plot compares the observed semi-major axis of the moon ($a_m$) with the theoretical Laplace limit (yellow shaded area), expressed as a function of the planet’s semi-major axis and mass relative to the star. 
    The coloured circles represent different exomoon candidates, with error bars showing reported uncertainties. 
    Candidates lying within the shaded region satisfy the Laplace stability condition, while those above it are unstable. 
    We find that all candidates are consistent with the Laplace criterion except for Kepler-1625b-i, which lies outside the stability region. 
    This is noteworthy since Kepler-1625b-i is likely the most studied exomoon candidate in the literature.}
    \label{fig:Laplace}
\end{figure}

Figure \ref{fig:Laplace} shows a representation of the Laplace criterion, by comparing the semi-major axis of moons ($a_m$) and the theoretical Laplace limit, which is represented by the yellow shaded area. We observe that all exomoon candidates lie within the stability region, except for one --- the candidate Kepler-1625-b-i, whose nature remains debated in the literature \citep{Teachey_2020}.

Recent work on WASP-49A b, a hot Saturn with atmospheric sodium potentially sourced from volcanic activity on an exomoon, suggests that stable moons may exist even in extreme close-in environments \citep{Sucerquia2025}. Using N-body simulations including planetary oblateness, relativistic precession, and tidal perturbations, they introduced a new stability metric (`selenity') and found that only Io-mass moons at $\sim 1.2$--$1.8\,R_{p}$ could survive, while smaller satellites are destabilised. 

\section{Observational detection limits}
\label{sec: observational_limits}

Exomoon detection relies primarily on indirect methods, since current telescopes cannot spatially resolve planet--moon systems. 
The most widely used approaches include TTVs and TDVs, which probe the dynamical perturbations introduced by a companion moon. 
In the TTV method, the barycentric motion of the planet–moon system causes periodic shifts in the observed mid-transit times, typically of the order of seconds to minutes depending on the moon's mass and orbital radius \citep{kipping_2009_transit1, kipping_2009_transit}. 
Transit duration variations arise as the orbital velocity of the planet around the barycentre modifies the transit chord (the path followed by the planet across the stellar disk) length across the star, leading to correlated variations in transit duration that are $\pi/2$ out of phase with the TTV signal \citep{kipping_2013_the}. 
A joint detection of both signatures is considered a strong indication of a satellite.

These estimates assume circular orbits and negligible mass ratios between the moon and planet, leading to first-order approximations for orbital periods and timing variations \citep{kipping_2009_transit1}:

\begin{equation}
P_p=2\pi \sqrt{\frac{a_p^3}{GM_*}}, \qquad P_m=2\pi \sqrt{\frac{a_m^3}{GM_p}},
\end{equation}
\begin{equation}
\delta_{\text{TTV}} \approx\frac{a_m}{V_p}\frac{M_m}{M_p}.
\end{equation}

Here, $V_p$ is the orbital velocity of the planet around the star, and $P_m$ and $P_p$ are the orbital periods of the moon and the planet, respectively.

Other methods such as gravitational microlensing can in principle detect exomoons through short-lived anomalies in the lensing light curve, with sensitivity peaking at planet–moon separations near the Einstein radius of the system (the characteristic scale of the lensing effect, depending on the masses and distances involved) \citep{kipping_2011_the}. 
Direct imaging and radial velocity techniques are theoretically viable for very massive moons or wide-orbit systems, but current sensitivity levels render them impractical except for extreme cases \citep{limbach_2021_on}. 
Additional ideas include pulsar timing, magnetic interactions (e.g. cyclotron radio emission), and machine-learning approaches such as convolutional neural networks trained on synthetic transit data \citep{weghs_2021_photometric, teachey_2021_identifying}, although none have yet produced confirmed detections.

While space missions such as Kepler and HST have provided key datasets, the limited time baseline and photometric precision have so far prevented robust exomoon confirmation. 
Future observatories such as PLATO and HWO are expected to significantly improve exomoon detectability by enhancing photometric sensitivity, time coverage, and the ability to discriminate between planetary and satellite signals \citep{limbach_2024_exomoons}.

\subsection{Limitations of current methods}
Despite significant progress in exoplanet detection, current observational techniques remain strongly limited in their sensitivity to exomoons, particularly for small satellites or compact planet--moon configurations.
Ground-based telescopes, under favourable conditions, might detect Earth-sized exomoons through TDV. However, these cases are rare and require high S/Ns and extensive monitoring \citep{teachey_2024_detecting}.

A key geometrical factor affecting detectability is the planet's impact parameter ($b$), which quantifies the projected distance (in units of stellar radii) between the transit chord, followed by the planet across the stellar disk. It is defined as

\begin{equation}
    b = \frac{a_p \cos i}{R_\star},
\end{equation}

where $i$ is the orbital inclination \citep{seager_2003_a}. Central transits correspond to $b = 0$, while $b \sim 1$ indicates grazing transits. Larger values of $b$ generally reduce the S/N of the transit and lead to less symmetric transit profiles because the planet crosses regions closer to the stellar limb, making exomoon detection via TTV or TDV more difficult. In our sample, inclination data are mostly unavailable or cluster near $i = 90^\circ$, limiting a detailed analysis of these effects.

Each detection method has its own set of theoretical limitations (\citet{kipping_2011_the}, \citet{kipping_2009_transit1}). For instance when using transits, small moons produce only slight dips in stellar brightness, which can be difficult to detect. Furthermore, favourable alignment is required, meaning the planet–moon system must transit across our line of sight. Other factors, such as stellar variability and instrumental noise, can obscure or mimic exomoon signals. On the other hand, although capable of detecting moons at wide orbital separations, microlensing events are rare and non-repeatable, making confirmation challenging. Additionally, the angular resolution is limited: the moon must be at a projected separation comparable to the Einstein radius to be detectable \citep{liebig_2010_detectability}. Moons too close to the planet blend with the planet's signal, while those farther away become undetectable \citep{liebig_2010_detectability}.

The upcoming Nancy Grace Roman Space Telescope will provide a unique opportunity to probe the exomoon population through gravitational microlensing. Simulations of its Galactic Exoplanet Survey \citep{Roman2025} indicate sensitivity down to $\sim 0.02\,M_{\oplus}$, comparable to Ganymede, for moons located at $0.1$--$0.5$ Hill radii around planets of $30\,M_{\oplus}$ to $10\,M_J$. Exomoons produce short-lived anomalies in the microlensing light curve, typically lasting a few hours, whose detectability depends critically on cadence and photometric precision. Although the nominal yield is expected to be low (of order one detection over the mission), Roman will establish the first statistical constraints on the frequency of massive exomoons, and may achieve higher detection rates under optimised follow-up strategies.

\begin{figure*}
    \centering
    \includegraphics[width=15cm]{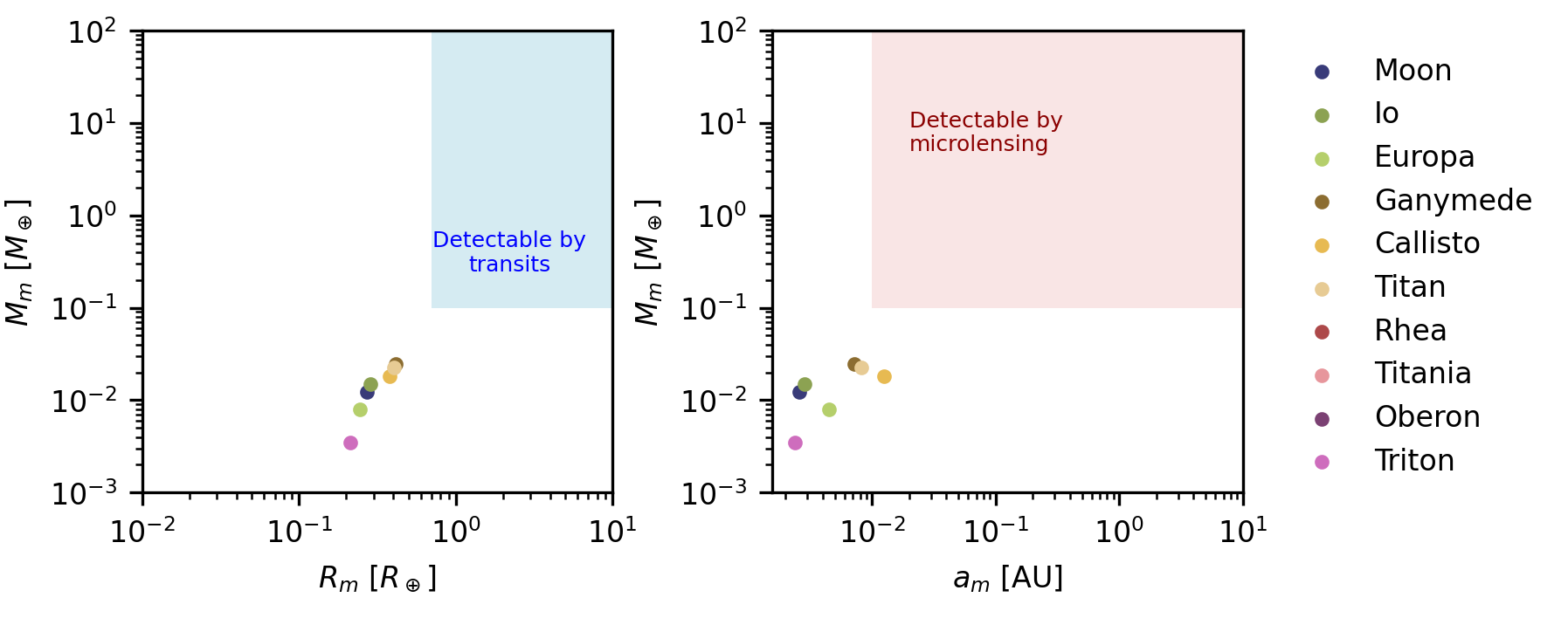} 
   \caption{
        Theoretical detection limits for exomoons using different observational techniques, 
        compared with major Solar System moons for reference (Table~\ref{tab:solarsystemmoons}). 
        Left: Detection space for the 'direct transit method', 
        where moons with radii $R_m \gtrsim 0.5\,R_\oplus$ are in principle detectable 
        through their own transit signal across the stellar disk (blue-shaded region; \citet{kipping_2009_transit1}). 
        This method differs from TTV and TDV techniques, which probe dynamical perturbations rather than the photometric dip of the moon itself. 
        Right: Sensitivity region for microlensing surveys (red-shaded region;  \citet{liebig_2010_detectability}). 
        Most Solar System moons lie below current detection thresholds, although future interferometric facilities may extend sensitivity towards Earth-mass moons at AU-scale separations.}
    \label{fig:methods}
\end{figure*}

Figure \ref{fig:methods} highlights the challenges in detecting smaller moons, using some of our Solar System moons for reference (see Table \ref{tab:solarsystemmoons}). The detection limits of current and future instruments may prevent the identification of smaller exomoons, indicating the need for improved detection methods to explore the full range of possible exomoon characteristics.

The Kepler mission's sensitivity to TTVs and TDVs also places constraints on detectability. The minimum detectable TTV is around 20–60 seconds, while TDVs might require larger amplitudes, typically 30–100 seconds (\citet{kipping_2009_on},\citet{Kipping:2012vt},\citet{kipping_2009_transit1}). For example, a moon with Earth-like mass and radius orbiting a Jupiter-sized planet at $20R_\mathrm{J}$ would induce a TTV of $\sim$105s and a TDV of $\sim$97s, well within Kepler’s detectability. In contrast, a Ganymede-like moon ($0.025M_\oplus$, $0.41R_\oplus$, $M_\oplus$ stands for the Earth's mass) would produce a TTV of $\sim$2.6s and a TDV of $\sim$2.4s, which fall well below detection thresholds. 

Recent JWST observations have provided some of the strongest empirical constraints on exomoon detectability. Deep searches with MIRI and NIRSpec time-series photometry have placed upper limits of approximately $R_m < 0.35\,R_\oplus$ and $M_m/M_p \lesssim 10^{-4}$ for short-period giants \citep{wilson2025deepsearchexomoonswise}. These limits are consistent with the theoretical detectability boundaries shown in Fig.~\ref{fig:methods}, supporting the view that current non-detections likely arise from instrumental sensitivity rather than the intrinsic absence of moons. Complementary analyses using dual-band Gaussian process modelling, where correlated noise is simultaneously fitted in two wavelength bands, demonstrate that JWST could recover transit depths as low as $\sim300$ ppm, corresponding to satellites with radii down to $\sim0.27\,R_\oplus$ (Io-sized) under favourable conditions \citep{householder2025sensitivity}.

While transit observations currently provide the strongest constraints, alternative techniques are also becoming increasingly relevant. Astrometry has recently emerged as a promising avenue for exomoon detection. The principle relies on measuring the reflex motion of a planet induced by an orbiting satellite, producing a signal with an angular amplitude:
\begin{equation}
\Delta \theta \simeq \frac{a_{m}}{d} \cdot \frac{M_{m}}{M_{p}},
\end{equation}
where $d$ is the distance to the system. 
Current VLTI/GRAVITY observations already reach the sensitivity to detect massive moons (e.g.\ $\sim 0.14\,M_J$ at 0.39~AU around AF~Lep~b; \citep{winterhalder_2026_astrometry}), while future facilities such as VLTI/PLANETES or the km-baseline interferometer (KBI) could probe down to Earth-mass satellites in habitable zones. Similarly, \citet{wagner2025astrometric} show that precise relative astrometry of directly imaged planets can detect Earth-mass moons around the giant planet candidate in the habitable zone of $\alpha$~Cen~A with the ELT, and even $\sim 0.2\,M_\oplus$ moons with a dedicated 3\,m class facility over a five-year baseline. Astrometry thus offers a complementary detection space to transit surveys, particularly for nearby systems where transit probabilities are low but astrometric precision can be maximised.

A recent VLTI/GRAVITY study reported tentative evidence of a massive exomoon candidate around the brown dwarf companion HD~206893~B, inferred from periodic astrometric deviations with a characteristic period of $\sim0.76$\,yr \citep{kral_2026_exomoon}. This result illustrates the growing potential of high-precision astrometry to probe wide-orbit moons around directly imaged planets and brown dwarfs, a parameter space largely inaccessible to transit-based techniques.

\subsection{Instrumental capabilities and detection thresholds}

Table \ref{tab:exomoon_detection} summarises the estimated detection thresholds for exomoons based on various current and future instruments. Radius limits are derived from transit-based methods (when applicable), while mass limits are estimated using conservative TTV detectability thresholds.

\renewcommand{\arraystretch}{1.2}
\begin{table*}
\caption{Estimated detection thresholds for exomoons with current and future observational facilities.}

\label{tab:exomoon_detection}
\centering
\renewcommand{\arraystretch}{1}
\begin{tabular}{lccccl}
\hline\hline
Instrument & $R_{\text{min}}$ ($R_\oplus$) & $M_{\text{min}}$ ($M_\oplus$) & Method(s) & Notes & Reference \\
\hline
HST    & $\sim$0.3      & $\sim$0.25     & Transits                  & Based on large gas giants & \citet{lammer_2014_origin} \\
JWST   & $\sim$0.2      & $\sim$0.05     & Transits, spectroscopy    & For favourable targets     & \citet{limbach_2021_on} \\
CHEOPS & 0.2--0.3       & $\sim$0.1--0.3 & Transits                  & Short baselines$^{(*)}$   & \citet{beck2017cheops} \\
PLATO  & $\sim$0.4      & 0.2--0.3       & Transits, TTV \& TDV         & $\delta_\text{TTV} \sim$ 30--60 s & \citet{Simon_2011, hippke2015detection} \\
HWO    & --             & $\sim$2$\times10^{-5}$ & Mutual events, TTVs     & $\delta_\text{TTV} \sim$ 10 s; no defined $R_{\text{min}}$ & \citet{limbach_2024_exomoons} \\
HEK    & $\sim$1        & $\sim$0.2      & Transits                  & Kepler-based programme      & \citet{kipping_2013_the} \\
\hline
\end{tabular}
\tablefoot{ This table lists the minimum detectable moon radius ($R_{\text{min}}$, in Earth radii $R_\oplus$), mass ($M_{\text{min}}$, in Earth masses $M_\oplus$), primary detection method(s), and relevant caveats under ideal observing conditions (bright host stars, multiple observed transits, and central transit geometry). References correspond to published limit estimates for each mission or programme. These thresholds provide estimates for the minimum detectable moon radius and mass under typical observing conditions. Mutual events refer to eclipses or occultations between planet and moon. $^{(*)}$\citet{Kalman2025} report practical CHEOPS limits closer to $0.6$--$0.8\,R_\oplus$ based on injection--retrieval experiments.}
\end{table*}

\begin{figure*}[ht]
    \centering
    \includegraphics[width=15cm]{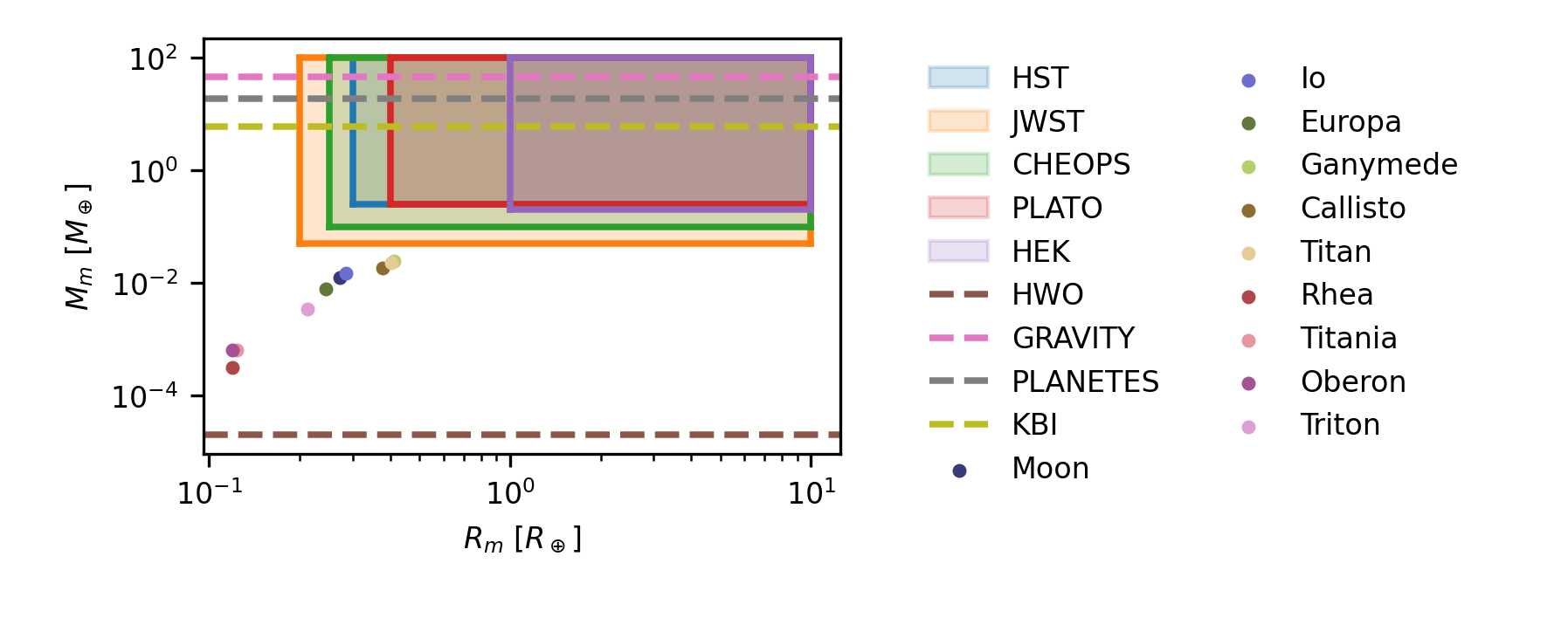}
\caption{
Detection limits of current and proposed facilities for exomoon searches, compared with the major moons of the Solar System (Table~\ref{tab:solarsystemmoons}). The shaded regions indicate the approximate parameter space accessible to transit and photometric techniques (HEK, HST, CHEOPS, JWST, and PLATO) in terms of moon radius ($R_m$) and mass ($M_m$), both expressed in Earth units. The horizontal dashed lines represent representative mass sensitivities for HWO and for astrometric facilities, including GRAVITY, its planned upgrade PLANETES, and a future KBI \citep{winterhalder_2026_astrometry}. Solar System moons are shown as coloured points for reference. Most known moons lie below current detection thresholds, illustrating the challenge of detecting sub-Earth-sized satellites. For the astrometric facilities, the quoted limits correspond to fiducial simulations assuming a 10\,$M_J$ planet at 20 pc and therefore should be regarded as illustrative rather than as universal detection thresholds. The figure highlights the complementarity of different detection techniques and the regions of parameter space most accessible to future exomoon surveys.
}
    \label{fig:limits_instruments} 
\end{figure*}

The detection limits of these instruments are illustrated in Fig.~\ref{fig:limits_instruments}. The shaded regions represent the detection windows based on the minimum detectable moon radius and mass.

For HWO, GRAVITY, PLANETES, and the proposed KBI, no reliable estimates of the minimum detectable moon radius are currently available, so only representative mass sensitivities are shown in Fig.~\ref{fig:limits_instruments}. For the astrometric facilities (GRAVITY, PLANETES, and KBI), these values are approximate and correspond to the fiducial configurations explored by \citet{winterhalder_2026_astrometry}, assuming a 10,$M_J$ planet located at 20 pc. Astrometric detectability depends on the moon-to-planet mass ratio, orbital separation, and system distance and therefore should not be interpreted as a universal mass threshold. Nevertheless, the comparison illustrates the approximate mass regime accessible to future facilities, with KBI potentially reaching the super-Earth regime ($\sim6,M_{\oplus}$). Without corresponding radius constraints, however, it remains uncertain whether such detections would also produce observable photometric or spectroscopic signatures.

None of the moons from our Solar System falls inside the confirmed detection regions of current or planned transit-based missions, which implies that, for the foreseeable future, only large and massive moons are likely to be detectable beyond the Solar System. Astrometric facilities provide a complementary pathway, since they are not limited by transit geometry and can probe nearby directly imaged systems, although their current sensitivity remains restricted to unusually massive satellites. A complementary approach has recently been proposed using radial-velocity monitoring of directly imaged substellar companions. Instead of detecting the moon itself, the method measures the reflex motion induced on the planetary-mass host. Applied to the brown dwarf companion CD-35 2722 B, \citep{Hoy_2026} reported evidence of one, and possibly two, massive satellite candidates. Although these candidates require independent confirmation, the study demonstrates that high-resolution spectroscopy may open a complementary region of exomoon parameter space beyond traditional transit searches.

Detecting a Callisto-like moon ($\sim$20 seconds TTV and TDV signal) requires missions with timing precision better than 20 seconds per transit, ideally over multiple transits. Assuming independent timing measurements with standard deviation \(\sigma_{\mathrm{transit}}\), the combined timing uncertainty after \(N\) transits decreases as
\begin{equation}
    \sigma_{\mathrm{total}} = \frac{\sigma_{\mathrm{transit}}}{\sqrt{N}}.
\end{equation}

To achieve a S/N of at least 3 for a signal \(\delta = 20\,s\), the condition is
\begin{equation}
\frac{\delta}{\sigma_{\mathrm{total}}} \geq 3 \Rightarrow N \geq \left(\frac{3 \sigma_{\mathrm{transit}}}{\delta}\right)^2.
\end{equation}

For \(\sigma_{\mathrm{transit}} = 20\,s\) and \(\delta = 20\,s\), this yields \(N \geq 9\) transits.

Exomoon detectability depends on whether the signal is expressed in terms of the satellite's radius (from the photometric transit depth) or its mass (from the timing variations induced on the host planet). 
Both approaches provide complementary constraints, and their sensitivity varies strongly with the observing strategy and instrumental precision.

\paragraph{Radius sensitivity}
While TTVs enable a direct inference of the satellite mass, transit photometry can constrain the minimum detectable exomoon radius. 
Simulations using the scatter-peak technique \citep{Simon_2011} suggest that PLATO could detect satellites as small as 
$R_m \approx 0.4\,R_\oplus$ around Jupiter-sized planets, assuming
\begin{itemize}
    \item  a photometric precision of $\sim50$ ppm,
    \item at least $\sim100$ observed transits, and
    \item planet orbital periods between 10–20 days.
\end{itemize}
This corresponds to ideal observing conditions: a bright host star ($V < 10$) and a gas giant with central transits. 
Such a radius threshold is comparable to Earth's Moon ($R_{\text{Moon}} \approx 0.27\,R_\oplus$), making PLATO potentially sensitive to super-lunar satellites. 

In contrast to transit-based missions, the proposed HWO is not expected to detect exomoons via photometric transit depth, but rather through mutual events such as transits, eclipses, shadows, and occultations \citep{cabrera2007detecting, limbach_2024_exomoons}. These events can produce substantial temporary flux variations (up to 30–40$\%$ for Earth–Moon analogues) and are most prominent at optical and near-infrared wavelengths (1.4–1.9\,$\mu$m), where exomoons may briefly outshine their host planets \citep{livengood2011properties}. Detecting such signals would require broadband photometric precision of $\sim$10\% over timescales of a few hours, with temporal cadences of $\sim$15 minutes and continuous monitoring spanning days to weeks. This method is especially promising for young, nearby systems ($d \lesssim 10$–20\,pc) hosting terrestrial planets at $\sim$1–5\,AU, where the expected frequency of mutual events is higher \citep{heller_2014_formation,limbach_2024_exomoons}. Recent studies indicate that HWO may be capable of detecting large habitable-zone exomoons through lunar-eclipse events under favourable observing conditions, potentially reaching sensitivities of $\sim0.5$--$0.9\,R_\oplus$ depending on the observing strategy \citep{limbach_2026_hwo}.

Polarimetric techniques provide a novel pathway for exomoon detection. Recent Monte Carlo radiative transfer simulations of Earth--Moon analogues show that mutual events, such as central transits and shadows, can induce polarisation variations up to $\Delta P \sim 2.7\%$ near quadrature \citep{Michaelis2025}. While such signals are beyond the sensitivity of current instruments, they could become detectable with next-generation facilities such as the HWO. Polarimetry thus offers a complementary diagnostic to flux-based methods, with the potential to constrain orbital inclinations and to disentangle planetary and lunar contributions to the observed signal.

\paragraph{Mass sensitivity}
The minimum detectable exomoon mass can be estimated through the amplitude of TTVs. 
For a single moon, the expected TTV amplitude is \citep{kipping_2009_transit1}
\begin{equation}
    \delta_{\text{TTV}} \approx \left( \frac{a_m}{a_p} \right) 
    \left( \frac{M_m}{M_p} \right) P_p ,
\end{equation}
which can be inverted to yield
\begin{equation}
    M_m^{\text{min}} \approx \frac{a_p}{a_m} \cdot \frac{M_p}{P_p} \cdot \delta_{\text{TTV}}^{\text{min}} .
\end{equation}

Applying this to different missions, we report the following results:
\begin{itemize}
    \item PLATO: For a Jupiter-like planet ($P_p = 10$\,days, $M_p = 318\,M_\oplus$, $a_p = 0.1$\,AU, $a_m = 0.01$\,AU), and $\delta_{\text{TTV}}^{\text{min}} = 60$\,s, we obtain
    \[
    M_m^{\text{min}} \approx 0.22\,M_\oplus .
    \]

    \item HWO: For an Earth-like planet ($P_p = 365$\,days, $M_p = 1\,M_\oplus$, $a_p = 1$\,AU, $a_m = 0.015$\,AU), and $\delta_{\text{TTV}}^{\text{min}} = 10$\,s, the estimate is
    \[
    M_m^{\text{min}} \approx 2.1 \times 10^{-5}\,M_\oplus .
    \]
\end{itemize}

These values highlight the complementarity of radius- and mass-based approaches. 
Transit methods constrain relatively large, super-lunar satellites, while TTVs offer the possibility to detect much smaller masses if timing precision and transit coverage are sufficient. 
Together, they define the practical parameter space accessible to current and future exomoon searches.

\subsection{Filtering strategy for exomoon candidates}
\label{sec: future_missions}

To identify favourable systems for exomoon detection, we applied a two-stage filtering process:

\begin{enumerate}
    \item observational filtering --- selecting systems observable with upcoming missions (PLATO and HWO) and
    \item dynamical filtering --- ensuring long-term orbital stability for potential moons.
\end{enumerate}

We began by downloading a catalogue of 6038 known exoplanets from the \texttt{exoplanet.eu} database \citep{martin_1995_encyclopaedia}. 
This catalogue includes all confirmed exoplanets with measured or estimated masses below the adopted upper limit of $60\,M_\mathrm{Jup}$, 
extended by the $1\sigma$ measurement uncertainty. 
This choice follows physical arguments suggesting a clear separation between giant planets and substellar objects at $\sim 60\,M_\mathrm{Jup}$. 
Earlier limits of $13\,M_\mathrm{Jup}$ (deuterium burning threshold) or $30\,M_\mathrm{Jup}$ (formation scenario) were therefore not applied here. 
As a result, the sample may contain objects in the brown dwarf regime, which were retained for completeness.

 The initial sample is restricted based on mission-specific observational limits. It focuses on targets accessible to the upcoming PLATO and HWO observatories.

\paragraph{PLATO constraints:}
We applied the following observational filters tailored to PLATO’s transit detection capabilities:
\begin{itemize}
    \item host stars with $4500\,\text{K} < T_\mathrm{eff} < 6500\,\text{K}$ to optimise asteroseismic precision, enabling accurate determinations of stellar masses, radii, and ages \citep{teachey_2024_detecting};
    \item planetary radii $R_p > 0.5\,R_J$ to enhance TTV and TDV signals \citep{kipping_2013_the};
    \item orbital periods $10\,\text{d} < P < 300\,\text{d}$ to ensure multiple observed transits over the mission lifetime \citep{dobos_2021_survival};
    \item transit impact parameters $b < 0.7$ to favour deeper, central transits \citep{kipping2010binning}; and
    \item visual magnitudes $V < 10$ to guarantee sufficient photometric precision \citep{rauer2016plato}.
\end{itemize}

\paragraph{HWO constraints:}
Given HWO’s capability for direct imaging via coronagraphy and reflected light observations, we adopted the following criteria to select candidate planets:

\begin{itemize}
    \item nearby bright host stars with distances $d < 20\,\text{pc}$ to maximise the S/N \citep{stark2019exoearth};
    \item planets with semi-major axes within the host star’s habitable zone, defined as
\begin{align}
a_\mathrm{HZ, inner} &= 0.95 \sqrt{\frac{L_\star}{L_\odot}} \quad \text{AU},\\
a_\mathrm{HZ, outer} &= 1.67 \sqrt{\frac{L_\star}{L_\odot}} \quad \text{AU},
\end{align}
    where the stellar luminosity (\(L_\star\)) is approximated by
    \begin{equation}
    \frac{L_\star}{L_\odot} = \left(\frac{R_\star}{R_\odot}\right)^2 \left(\frac{T_\mathrm{eff}}{T_\odot}\right)^4,
    \end{equation}
    with \(T_\odot = 5772\,\text{K}\), \(R_\odot\) being the solar radius, and $L_\odot$ the solar luminosity \citep{Kopparapu_2013};
    \item Earth-like planetary radii in an extended range $0.5\,R_\oplus \lesssim R_p \lesssim 1.5\,R_\oplus$ ($\sim 0.044 - 0.133\,R_J$) to include a broader set of terrestrial candidates;
    \item visual magnitudes $V < 10$ for adequate imaging sensitivity \citep{stark2019exoearth};
    \item host star effective temperatures constrained to $4500\,\text{K} < T_\mathrm{eff} < 6500\,\text{K}$ to select Sun-like stars optimal for detection \citep{stark2019exoearth};
    \item preference for young systems (ages approximately less than or equal to a few hundred megayears) where the frequency of mutual events is expected to be higher \citep{heller_2014_formation, limbach_2024_exomoons}; and
    \item planets with semi-major axes of $\sim$1–5 AU, where moons are most likely to be present and detectable in reflected light \citep{heller_2014_formation}.
\end{itemize}

Applying the PLATO observational criteria yields 610 candidate exoplanets, while the HWO criteria yields 14 candidates. The full lists of PLATO and HWO candidates are provided as supplementary material available at the \href{https://github.com/Aliper02/Exomoons_Catalogue.git}{Github repository}.

As described in Sect.~\ref{sec:stability_criteria}, we applied three dynamical criteria to assess the orbital stability of potential exomoons: Hill, Roche, and Laplace limits. These criteria define the dynamical boundaries within which a moon can remain gravitationally bound to its host planet over long timescales, avoid tidal disruption, and resist destabilising stellar perturbations.

To implement these filters on our observationally selected planet samples, we developed a Python-based analysis pipeline with the following assumptions and methodology:

\begin{itemize}
    \item Moon masses: We tested a grid of assumed moon masses spanning from $0.0005\,M_J$ (approximately $0.16\,M_\oplus$ or 2 $M_\mathrm{Moon}$) up to $0.05\,M_J$ (about 15 $M_\oplus$) to cover a range of plausible exomoon sizes.
    \item Orbital eccentricities: Moon eccentricities, $e_m$, of 0.0, 0.1, and 0.3 were considered to explore their effect on stability limits. When planet eccentricity, $e_p$, is unavailable, it is assumed to be zero.
    \item Planet and star parameters: Planet mass ($M_p$) and radius ($R_p$) were taken from the catalogues in Jupiter units, semi-major axis ($a_p$) in astronomical units (AU), and stellar mass ($M_*$) in solar masses, converted internally to Jupiter masses where needed for formula consistency.
\end{itemize}

For each planet, the code iterates over the moon mass and eccentricity grid to determine whether there exists a moon orbital semi-major axis satisfying all three criteria simultaneously. The moon mass and eccentricity combination yielding stable solutions is recorded for diagnostics.
The full Python code used to perform the observational and dynamical filtering 
is publicly available at our \href{https://github.com/Aliper02/Exomoons_Catalogue.git}{Github repository}.

Applying the methodology from the dynamical filtering yields the following.

\begin{itemize}
    \item HWO candidates: None of the 14 planets observable with the Habitable Worlds Observatory meets any of the stability criteria for any tested moon mass or eccentricity combination, implying that detectable stable exomoons are unlikely in this sample under current assumptions.
    \item PLATO candidates: Out of 610 planets meeting observational filters for the PLATO mission, 27 satisfy all orbital stability criteria and are flagged as promising hosts for dynamically stable exomoons. The detailed list of these stable candidates is provided in the \href{https://github.com/Aliper02/Exomoons_Catalogue.git}{Github repository}.
\end{itemize}
These results underscore the utility of dynamical filtering to refine exomoon search targets and suggest that the PLATO mission's transit survey may offer a more favourable sample for discovering stable exomoons given our adopted assumptions. In contrast, the lack of viable HWO candidates in our filtered sample should not be interpreted as a fundamental limitation of the mission itself. Since HWO will primarily discover new planets via direct imaging around nearby stars, its most promising targets are not yet part of current exoplanet catalogues and thus fall outside the scope of our present analysis.

Our method builds upon frameworks from recent literature. For example, \citet{makarov_2023_pathways} applied both Hill and Roche stability criteria to known exoplanets to identify favourable candidates for exomoon detection via transits. Their results support our filtering approach, highlighting gas giants with periods of 50–300 days as optimal.

{\subsection{Stable exomoons in the PLATO FOV}

To determine which stable candidates could be observed by PLATO, we considered the official Long Observation Phases (LOPs) defined for the mission \citep{PLATO_2017}. 
PLATO will not observe the entire sky simultaneously but instead monitor selected long-duration fields for several years in order to maximise transit detections and asteroseismic characterisation \citep{Rauer_2014}. 

Following the current mission design, we include both the southern long-duration field (LOPS2) and the northern long-duration field (LOPN1), as discussed in recent PLATO observing strategy studies \citep{Nascimbeni2022, Nascimbeni2025}. 
For simplicity, each field is approximated as a $49^\circ \times 49^\circ$ rectangular region in equatorial coordinates, consistent with the effective sky coverage of the PLATO camera mosaic.

The adopted field centres are
\begin{itemize}
    \item LOPS2: centred at $\mathrm{RA}=95.31^\circ$, $\mathrm{Dec}=-47.89^\circ$
    \item LOPN1: centred at $\mathrm{RA}=277.18^\circ$, $\mathrm{Dec}=52.86^\circ$.
\end{itemize}

Although the real PLATO footprint has an irregular geometry due to the overlap of the 24 cameras, the rectangular approximation provides a practical and reproducible criterion for catalogue-based candidate selection \citep{Rauer_2014, Nascimbeni2022}.

We applied this approximation to our catalogue of stable candidates by identifying those located within these regions. This allowed us to rigorously determine which systems are accessible during PLATO's nominal observing strategy, noting that each long-duration field remains fixed for several years before potential re-pointing \citep{Rauer_2014}.
\begin{figure}
    \centering
    \includegraphics[width=\columnwidth]{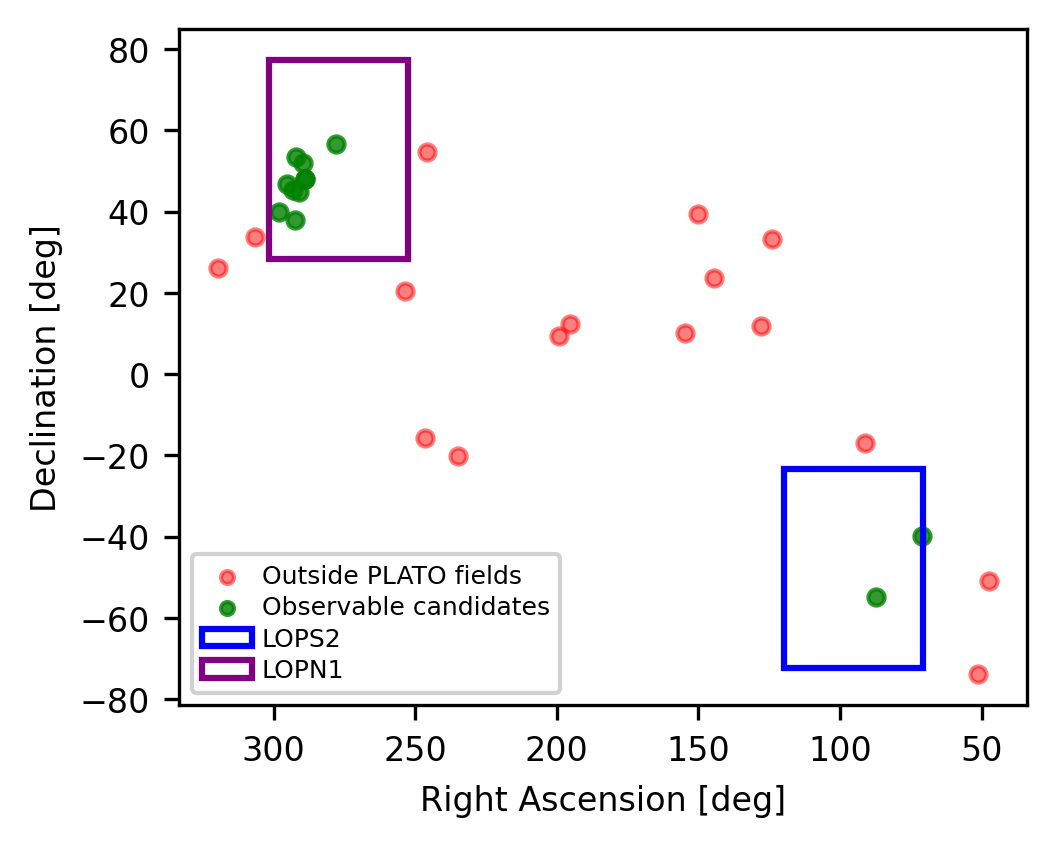}
    \caption{Distribution of dynamically stable exomoon candidates in equatorial coordinates (RA, Dec). 
    The blue and purple rectangles indicate the southern (LOPS2) and northern (LOPN1) PLATO long-observation fields, repectively, both approximated as $49^\circ \times 49^\circ$ regions centred on the official PLATO pointing coordinates \citep{Nascimbeni2025, Nascimbeni2022}. Stable candidates located within either field are shown in green, while those outside both fields are shown in red.}
    \label{fig:plato_visibility}
\end{figure}
The results show that a total of 12 stable candidates fall within either LOPS2 or LOPN1 (Fig.~\ref{fig:plato_visibility}). 
In the southern field (LOPS2), we identify two systems: NGTS-29~b ($\mathrm{RA}=70.9975^\circ$, $\mathrm{Dec}=-39.9067^\circ$) and TOI-201~b ($\mathrm{RA}=87.4000^\circ$, $\mathrm{Dec}=-54.9108^\circ$) \citep{Nascimbeni2025}. 

In the northern field (LOPN1), we find ten additional candidates: Kepler-117~b/c, Kepler-1514~b, Kepler-1657~b, Kepler-420~Ab, Kepler-47~(AB)d, Kepler-64~(AB)b, Kepler-86~b, TOI-2010~Ab, and TOI-2180~b \citep{Nascimbeni2022}.
Interestingly, although LOPS2 represents the most established long-duration pointing in the current mission baseline, the LOPN1 configuration encompasses a significantly larger number of dynamically stable candidates in our catalogue. This makes it particularly attractive for future exomoon studies, subject to the final mission implementation.

Additionally, we explored an optimised placement of a hypothetical $49^\circ \times 49^\circ$ field maximising the number of stable candidates. 
The optimal pointing is found near $\mathrm{RA}=295.30^\circ$, $\mathrm{Dec}=46.92^\circ$, which would include up to 12 stable candidates simultaneously.
All catalogues generated in this work, including stability classifications and sky coordinates, are publicly available in the associated GitHub repository:
\href{https://github.com/Aliper02/Exomoons_Catalogue.git}{Exomoons Catalogue}.

\section{Other disruption mechanisms and future work}
\label{sec:disruption}

\subsection{Planetary migration, resonances, and moon loss}

Planetary migration, a fundamental process in planet formation, can strongly affect the stability and survival of satellite systems. Planets may undergo different migration regimes—types I, II, or III—depending on their mass and interaction with the protoplanetary disk \citep{marov_2021_migration}. Inward migration, especially of giant planets, can destabilise moons through gravitational and tidal forces.

A particularly disruptive process is the evection resonance, which occurs when the moon's pericentre precession rate matches the planet’s orbital frequency. This can increase the moon's eccentricity beyond stability thresholds, leading to collision, tidal disruption, or ejection \citep{spalding_2016_resonant}. The condition for resonance is

\begin{equation}
    J_2 \frac{R_p^2}{a_m^2} n_m = n_p,
\end{equation}

where $J_2$ is the planet’s quadrupole moment and $n_m$, $n_p$ are the mean motions of the moon and planet, respectively. Simulations suggest that hot Jupiters may have lost any primordial moons during migration \citep{makarov_2023_pathways}.

The conditions under which planets form also influence the likelihood of forming and retaining moons. In particular, \citet{nesvorny2025terrestrial} show that the dynamical environment of planet formation plays a key role in determining the potential for satellite formation and long-term retention. The structure of the protoplanetary disk, planetary migration, and early dynamical instabilities directly impact the Hill stability and accretional environment required for large moons to form and persist. Systems with tightly spaced terrestrial planets may exhibit long-term orbital stability, yet often lack the late giant impacts typically associated with moon formation. In the context of exomoons, we note the following:

\begin{itemize}
\item Rapidly formed and isolated planets, such as Mars, are unlikely to host massive moons because of the short lifetime of circumplanetary material.
\item Planetary migration can mix material originating from different regions of the protoplanetary disk, creating compositional gradients that may hinder the giant-impact scenarios thought to produce Earth–Moon-like systems.
\item Giant planet migration may suppress exomoon formation unless migration ends before satellite accretion or the system remains dynamically quiescent over long timescales.
\end{itemize}

Circumbinary planets constitute a particularly hostile environment for satellite retention. $N$-body simulations by \citet{Gordon2025} indicate that during planetary migration most moons either become unbound and evolve into independent 'ploonets' (former exomoons that become gravitationally detached from their host planet) or are ejected entirely, with only $\sim 15$--$30\%$ surviving as true satellites. The critical parameter is the ratio $\gamma = a_m / r_H$, with moons remaining bound only if $\gamma < 0.48$. Importantly, a non-negligible fraction of surviving moons and ploonets lie in the habitable zone of their host binaries, while complete ejections may contribute to the observed population of free-floating planetary-mass objects. These results suggest that circumbinary planets may be less favourable targets for exomoon searches despite their long orbital periods.

Numerical simulations of exomoon survival during hot Jupiter migration show that retrograde configurations are significantly stabler than prograde ones. In disk-driven migration scenarios, retrograde moons survive in up to $\sim20\%$ of cases, whereas prograde survival fractions remain below $5\%$, with stability largely independent of the satellite's mass \citep{Pu2025}. In contrast, secular coplanar high-eccentricity migration is far more destructive: only close-in, massive retrograde moons ($M \gtrsim 0.01\,M_J$, $a_m \sim 2$--$3\,R_J$) can persist, with survival probabilities of order 1\%. Interestingly, a few percent of free-floating planets produced by scattering retain their moons, suggesting that planet–moon pairs may exist in interstellar space. These results reinforce the expectation that hot Jupiters are unlikely to host detectable prograde moons, and that retrograde survivors, if any, are rare and massive.

The dynamics of multi-planet resonant systems also bear directly on the long-term stability of potential exomoons. Recent $N$-body simulations of the TOI-178 system \citep{Boskovic2025} show that resonant planetary architectures can strongly destabilise small bodies and potential exomoons. Co-orbital clearing regions may be 30--50\% larger than predicted by classical theory, while resonance chains rapidly generate Kirkwood-like gaps (depleted regions produced by resonant gravitational interactions) on timescales of a few hundred years. In addition, resonance-driven inclination oscillations can dynamically excite satellites and reduce their long-term stability. These results suggest that compact resonant systems may be intrinsically unfavourable environments for long-lived exomoons.

\subsection{External perturbations and dynamical instability}

Exomoons may also be destabilised by gravitational perturbations from external bodies (x). Perturbations are most effective when
\begin{itemize}
  \item the distance of the encounter satisfies \( d_x \leq 3 r_H \);
  \item the mass ratio between the perturber and planet exceeds \( M_x / M_p \gtrsim 0.01 \);
  \item the moon has low mass (\( M_m / M_p \ll 1 \)) or is distant;
  \item the orbit of the perturber is coplanar; and
  \item the encounter velocity is low enough to enable significant gravitational interaction.
\end{itemize}
These criteria are derived from force-based approximations and supported by three-body dynamics \citep{heggie_2025_the}.

Following \citet{heggie_2025_the}, we adopted a perturbation efficiency parameter defined as
\begin{equation}
\label{epsilon}
    \epsilon = \frac{F_{x-m}}{F_{p-m}} = \left( \frac{a_m}{d_x} \right)^2 \frac{M_x}{M_p},
\end{equation}
which quantifies the relative strength of the gravitational perturbation from an external body on a moon compared to the host planet. When $\epsilon > 1$, the moon is likely to become unstable. 

Figure \ref{fig:stability_diagram} illustrates the regions of stability as a function of perturber mass and distance, for a nominal Earth–moon-like system.

\begin{figure}
    \centering
    \includegraphics[width=\hsize]{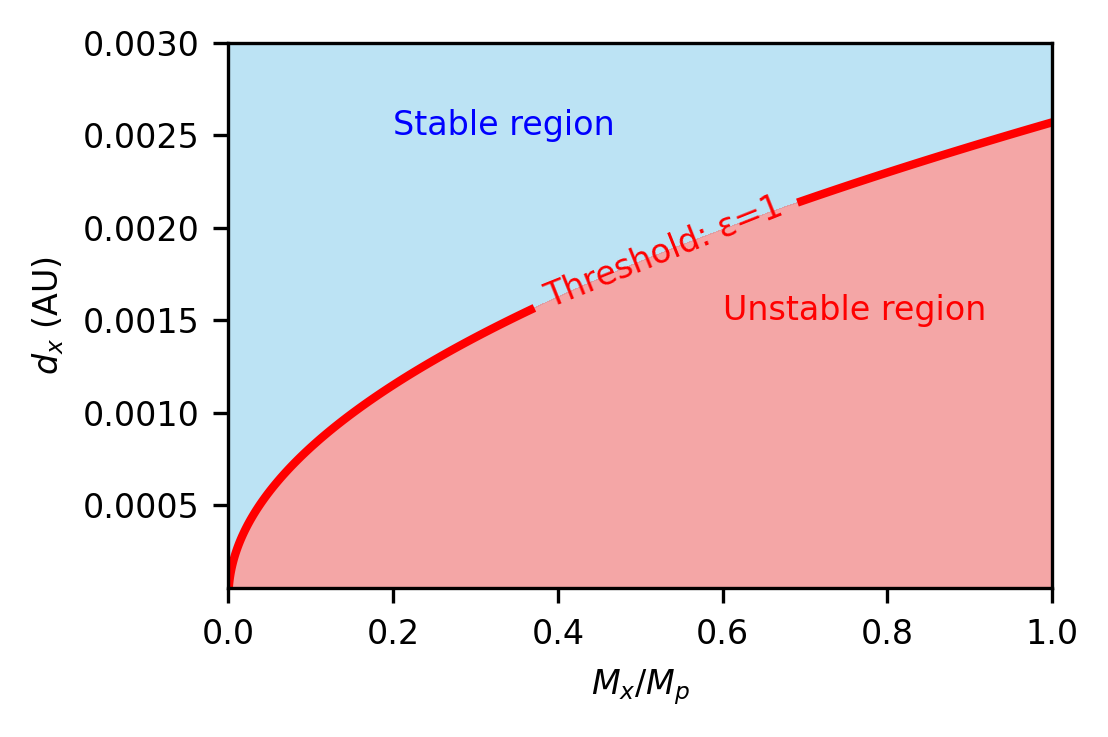}
    \caption{ Stability diagram for exomoon systems as a function of the perturber-to-planet mass ratio ($M_x/M_p$) and encounter distance ($d_x$). 
The red curve marks the critical threshold $\epsilon = 1$, separating unstable configurations ($\epsilon > 1$, red region) from stable configurations ($\epsilon < 1$, blue region). Moons below the critical boundary are expected to undergo dynamical disruption due to external perturbations, whereas systems above the boundary may remain stable over long timescales.}
    \label{fig:stability_diagram}
\end{figure}

To illustrate the applicability of Eq. \ref{epsilon}, we consider the effect of the most massive non-planetary bodies in the Solar System on existing moons. The dwarf planet Ceres ($M_x \approx 9.4 \times 10^{20}$ kg) and the large asteroid Vesta ($M_x \approx 2.6 \times 10^{20}$ kg) represent the strongest potential perturbers among minor bodies. For comparison, the Earth has a mass $M_p = 6.0 \times 10^{24}$ kg, and Mars has a mass $M_p = 6.4 \times 10^{23}$ kg \citep{russell2011dawn}.

Applying the critical condition $\epsilon = 1$, the distance at which an external perturber would significantly destabilise a moon is given by

\begin{equation}
    d_x = a_m \sqrt{\frac{M_x}{M_p}}.
\end{equation}

For the Earth–Moon system ($a_m = 3.84 \times 10^5$ km), Ceres would need to approach the Moon within
\[
d_x \approx 4.6 \times 10^3 \, \text{km},
\]
comparable to the lunar radius. Such an encounter is effectively impossible under present Solar System conditions, showing that even the most massive minor bodies cannot destabilise the Earth–Moon configuration.

In contrast, for Mars and its small satellite Phobos ($a_m = 9377$ km), the same calculation yields
\[
d_x \approx 950 \, \text{km}.
\]
Thus, in principle, a close flyby of a Ceres-mass object at a distance of the order of $10^3$ km from Phobos could disrupt its orbit. A similar calculation with Vesta yields a critical distance of $\sim 600$ km. While such close approaches are extremely improbable in the present Solar System, these results demonstrate that small moons orbiting low-mass planets are substantially more vulnerable to external perturbations.

These examples emphasise that gravitational perturbations from large asteroids or dwarf planets are dynamically irrelevant for massive, tightly bound moons such as the Earth’s Moon. They could, however, play a role in destabilising small moons in other systems, particularly in dense exoplanetary environments where close encounters are more likely.

\subsection{Energy-based criteria and future work}

Energy-based analysis provides complementary insight. If the total energy of the moon relative to the system becomes positive due to a perturbation, it may escape:

\begin{equation}
E_{\text{total}} = \frac{1}{2}M_m v^2 - \left( \frac{GM_p M_m}{a_m} + \frac{GM_\star M_m}{d_{\star-m}} + \frac{GM_x M_m}{d_x} \right).
\end{equation}

where $G$ is the gravitational constant, $v$ is the moon's relative velocity, and $d_{\star-m}$ is the distance between the star and the moon.
Alternatively, the evolution of the Jacobi constant in the three-body problem can signal long-term instability \citep{mary_solar}.

In this work, we used analytical criteria to estimate perturbation effects and illustrate regions of likely disruption. Future studies will benefit from $N$-body numerical simulations using tools such as \texttt{REBOUND} \citep{rein_2012_rebound}, which allow the modelling of orbital evolution under complex perturbative scenarios. These simulations are essential for assessing long-term survivability of moons in dynamically active systems. Tidal evolution may severely limit the survival of moons around short-period habitable-zone planets.
Tidal evolution may further restrict exomoon survival around short-period habitable-zone planets. Simulations of the Hycean candidate K2-18b using the \texttt{reboundx} tidal module suggest that lunar-mass satellites become unstable within $\lesssim10$ Myr under both Earth-like and Neptune-like tidal parameters \citep{Patel2025}. The rapid outward migration of the moon driven by the planet's spin and small Hill sphere leads to ejection long before the system age ($\sim 3$ Gyr), effectively ruling out observable moons. These results suggest that habitable-zone planets around M dwarfs are unlikely to retain long-lived moons.

Recent long-term $N$-body simulations further demonstrate that analytical stability limits may underestimate the complexity of exomoon dynamics. 
\citet{barbosa_2026_on} derived updated empirical relations for the outer stability boundary of prograde exomoons as a function of satellite eccentricity and planet-to-star mass ratio, obtaining:
\begin{equation}
\frac{a_E}{R_{H}} \simeq 0.48781 - 0.12753\,e_{\rm m},
\end{equation}
where $a_E$ is the critical semi-major axis for long-term stability. Their simulations reveal a strong dependence on the initial orbital phase of the satellite, with moons initialised near opposition surviving at larger fractions of the Hill radius than predicted by classical prescriptions. 
They also identify detached `stability islands' associated with evection resonances, particularly for large planet-to-star mass ratios, indicating that resonant phase protection can maintain long-lived moons beyond simple analytical limits. 
These results suggest that future exomoon filtering pipelines should combine analytical stability criteria with dedicated $N$-body simulations to more accurately constrain the parameter space of dynamically stable satellites. Consequently, some systems rejected by simplified analytical filters may still permit long-lived exomoons under favourable resonant or phase-protected configurations.

Beyond bound planetary systems, some studies have considered the fate of planet--moon pairs after dynamical ejection. Atmospheric models by \citet{Dahlbudding2026} suggest that exomoons orbiting free-floating planets could maintain habitable surface conditions through tidal heating and thick H$_2$-dominated atmospheres, even in the absence of stellar irradiation \citep{Dahlbudding2026}. These studies indicate that collision-induced absorption by H$_2$ can sustain surface temperatures above the freezing point of water for gigayear timescales, expanding the potential habitability of free-floating planet--moon systems produced during dynamical evolution.

\section{Comparative analysis with the Solar System}
\label{sec:comparative}

The study of exomoons gains valuable context when contrasted with the diverse and well-characterised moons in our own Solar System. These natural satellites provide useful analogues in terms of size and composition, helping inform both the selection criteria and the expected characteristics of exomoon candidates \citep{barr_2016_formation}. Table \ref{tab:solarsystemmoons} (in Sect. \ref{app:ssmoons}) lists some of the major moons from the Solar System. 

The largest gas giants in the Solar System, Jupiter and Saturn, are accompanied by complex satellite systems. Jupiter, the most massive planet, hosts four large moons — Io, Europa, Ganymede, and Callisto — with orbital distances ranging from 6 to 26 Jupiter radii ($R_J$). These satellites span a wide range of sizes and compositions. For example, Ganymede, at $0.025M_\oplus$ and 5,268 km in diameter, is the most massive moon in the Solar System, while Io is the most volcanically active body, driven by tidal heating from orbital resonances \citep{showman1997coupled, schubert2004interior}. Saturn’s satellite system is dominated by Titan, a $0.022M_\oplus$ body orbiting at 20 Saturn radii ($R_S$), unique for its thick nitrogen-rich atmosphere and surface hydrocarbon lakes \citep{lunine2009saturnstitanstricttest, lopes2019titan}. Gas giant moons can reach masses comparable to Mercury and occupy dynamically stable, long-lived orbits, making them relevant analogues for potential exomoons. The disparity in the number of large satellites between these two planets is thought to result from differences in the timing and scale of gas infall onto their circumplanetary disks. Jupiter may have been particularly efficient at carving a deep gap in the protoplanetary disk, thereby isolating a reservoir of solid material beyond its orbit that could later be accreted into the circumplanetary disk and incorporated into large satellites. 
Hydrodynamical simulations demonstrate that a Jupiter-mass planet can indeed open an annular gap in the disk, significantly reducing the gas surface density near its orbit \citep{1996ApJ...460..832T}.  
Moreover, models by \cite{ronnet2020formation} show that this gap facilitates the accumulation of planetesimals at its outer edge, providing the building blocks required for the formation of the Galilean satellites.  

To contextualise the mass ratios observed in candidate exomoons, we compare them to those found in moons of the Solar System. The ratio $M_m/M_p$ provides a dimensionless parameter that helps assess how gravitationally dominant a planet is with respect to its satellite. In Fig. \ref{fig:histo} both exomoon candidates and Solar System moons are shown, revealing a clear difference in mass ratios despite the limited sample size. If we base our expectations solely on the moons of the Solar System, the existence of large exomoons may appear unlikely.

\begin{figure}
    \centering
    \includegraphics[width=\hsize]{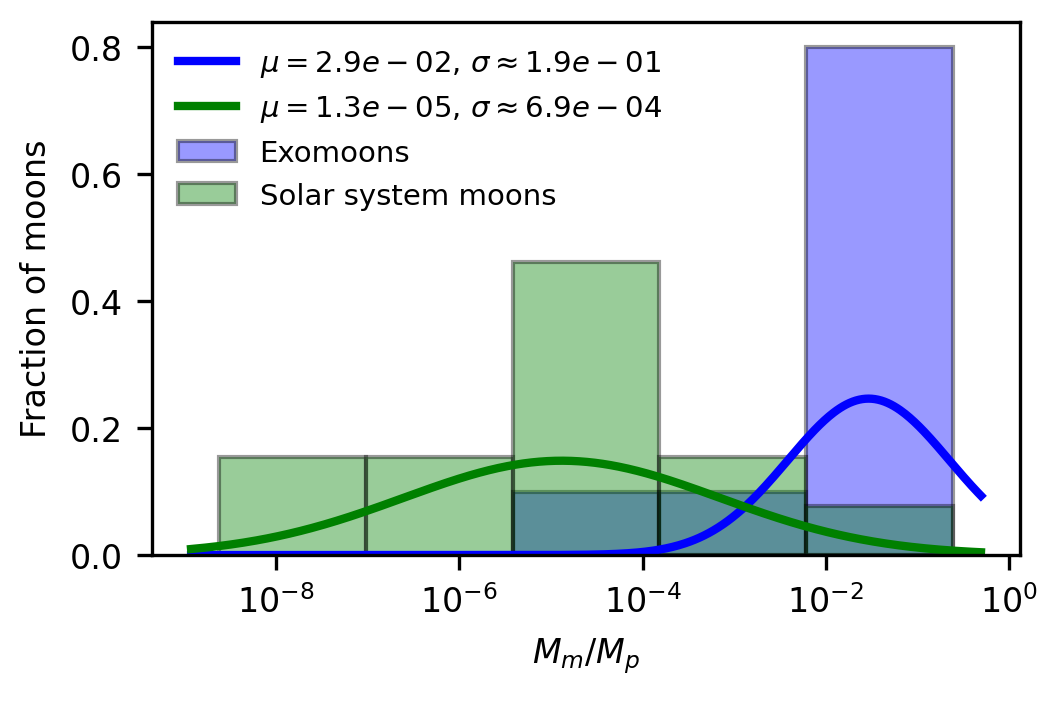}
     \caption{Normalised distribution of the moon-to-planet mass ratio ($M_m/M_p$) for Solar System moons (green) and exomoon candidates (blue), together with Gaussian fits (solid lines). The histogram highlights the different regimes of mass ratios: Solar System moons are typically found at low values ($M_m/M_p \sim 10^{-7} - 10^{-4}$), whereas the reported exomoon candidates are shifted towards much higher values, up to $M_m/M_p \sim 10^{-1}$. The Gaussian fits yield $\mu_\mathrm{exo} = -1.53$, $\sigma_\mathrm{exo} = 0.88$ for exomoon candidates, and $\mu_\mathrm{SS} = -4.88$, $\sigma_\mathrm{SS} = 1.73$ for Solar System moons, showing quantitatively the systematic shift between the two populations.}
    \label{fig:histo}
\end{figure}

As shown in Fig. \ref{fig:histo}, 80\% of candidate exomoons have mass ratios ranging from $10^{-2}$ to 1, while 50\% of the selected Solar System moons have mass ratios close to $10^{-5}$. The Gaussian fits indicate that the distributions are significantly different, with exomoon candidates having $\mu_\mathrm{exo} = -1.53$, $\sigma_\mathrm{exo} = 0.88$, and Solar System moons $\mu_\mathrm{SS} = -4.88$, $\sigma_\mathrm{SS} = 1.73$. 

If we examine the mass ratio $M_m / M_p$ versus the semi-major axis $a_m$ (Fig. \ref{fig:exomoons ratio}), we can see that exomoon candidates tend to exhibit larger moon-to-planet mass ratios and wider orbital separations than the moons of the Solar System. This is consistent with current observational limitations for exomoon detection, which currently restrict detections to large (and thus massive) satellites. In addition, indirect methods such as TTV and TDVs require moons with a sufficiently wide orbital separation from their host planet to produce a detectable dynamical signal, while at the same time favouring short-period planets so that multiple transits can be observed within the mission lifetime.

Figure \ref{fig:exomoons ratio} is particularly interesting because it highlights that Earth's Moon is the only satellite in the Solar System with a mass ratio exceeding \(M_m/M_p \approx 0.01\). In contrast, most other moons in the Solar System, including even such large ones as Ganymede, Titan, and Triton, have much smaller mass ratios relative to their host planets, meaning their tidal effects are less pronounced. This difference underscores the unique dynamical relationship between Earth and its Moon, which is likely a key factor in supporting the planet's habitability over geological timescales.

\subsection{Formation of moons}
\label{subsec:formation}
Exomoon candidates suggest diverse formation scenarios depending on their properties. While co-formation is typically associated with mass ratios of \( M_m / M_p \sim 10^{-4} \), many of the exomoon candidates exhibit significantly higher values, up to \( \sim 0.1 \) in some cases, making alternative formation mechanisms more likely. For instance, Kepler-1513 b-i, KOI-268.01-i, WASP-49b-i, and WASP-76 Ab-i fall within or near the expected range for co-formation and could be consistent with this process, especially at moderate distances from their hosts. However, most candidates, including Kepler-1625 b-i, Kepler-517Ab-i, and Kepler-1000b-i, have inferred mass ratios \( M_m / M_p \gtrsim 10^{-2} \), suggesting capture or giant impact scenarios instead. Kepler-1625 b-i, in particular, fits the capture hypothesis due to its high mass and possible orbital misalignment. Other systems, such as MOA-2011-BLG-262L b-i, MOA-2015-BLG-337a-i, and 2MASS J2117-2940-i, might also result from capture, though detailed orbital data are lacking.
Kepler-409b-i and Kepler-1442b-i, whose host planets are thought to be rocky, could have formed through giant impacts instead.

Theoretical models of circumplanetary disks indicate that giant planets can form systems of massive moons with a total mass scaling as $M_{\mathrm{sat}} \sim 10^{-4} M_p$, in agreement with Solar System constraints \citep{Dencs2025}. However, planetary migration acts as a destructive mechanism: outer satellites are stripped during inward migration, leaving only close-in survivors, a process dubbed `grand theft moons'. This scenario explains the paucity of exomoons around hot and warm giants and implies that habitable exomoons are most likely to be found around giant planets that remained on wide orbits. Given that close-in hot Jupiters are unlikely to retain long-term stable moons, any such satellites may be transient captures or artefacts of complex system dynamics (e.g. HD 189733 Ab-i, WASP-12b-i).

In addition, dynamical evolution during and after formation plays a critical role in determining long-term moon survival. Efficient damping of orbital eccentricity may favour the formation of dynamically stable planetary architectures and enhance the long-term survival prospects of satellites \citep{nesvorny2025terrestrial}. Systems where eccentricity damping was inefficient are expected to experience stronger tidal interactions and secular perturbations, leading to orbital instability or ejection of satellites.

\subsection{Stability of moons} 
Regarding stability, all the moons in the Solar System exhibit very low eccentricities, remaining well below the upper limits set by both the Hill and Roche stability criteria as seen in Fig. \ref{fig:SS stb}, in contrast with the exomoon case (Fig. \ref{fig:stability criteria}). This confirms that their orbits are dynamically stable and that they are not currently at risk of tidal disruption or gravitational destabilisation. Furthermore, the Hill limit values not shown in the figure all lie above 1, which implies that any value of eccentricity would satisfy this criterion for those moons, thus highlighting their strong dynamical stability. Although not shown in the figure, all Solar System moons also fulfil the Laplace stability criterion, reinforcing the robustness of their orbital configurations.

\subsection{Host star properties}

Host star properties can influence both the long-term stability of exomoon systems and their observational detectability. Figure~\ref{fig:edad_T} lists the effective temperatures and estimated ages, when available, of host stars associated with candidate exomoons listed in our \href{https://github.com/Aliper02/Exomoons_Catalogue.git}{Github repository}.

\begin{figure}
    \centering
    \includegraphics[width=\hsize]{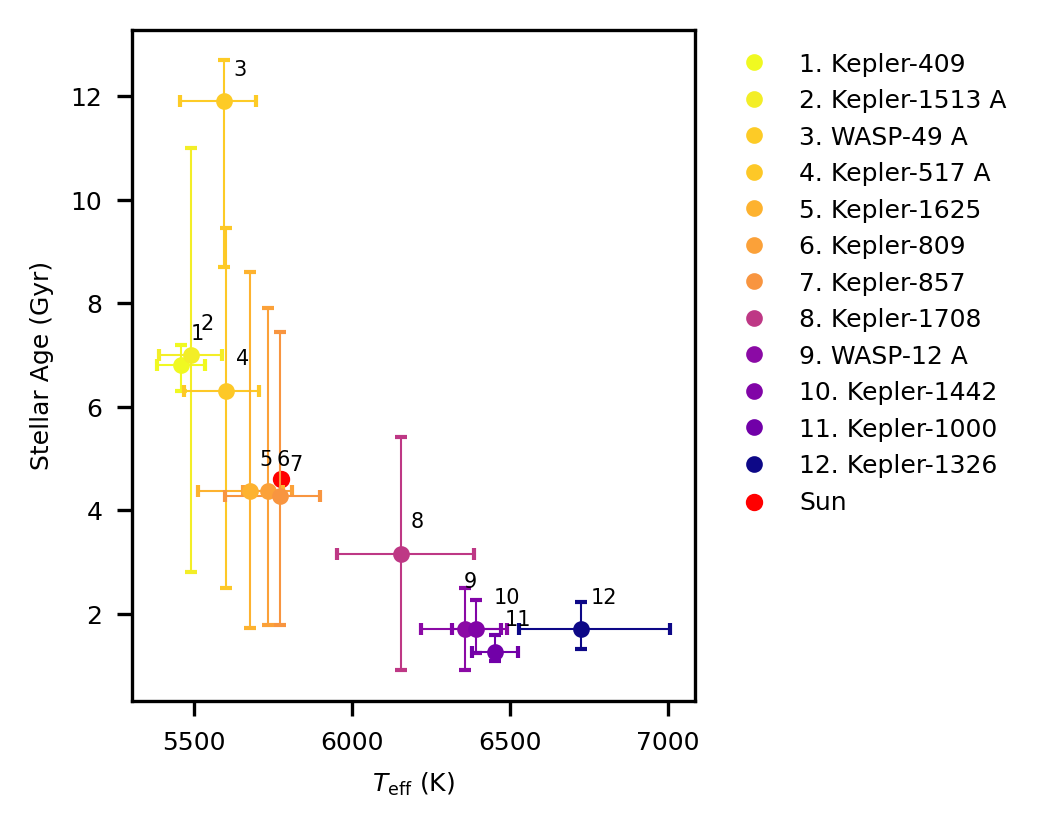}
    \caption{
        Effective temperature and estimated stellar ages of host stars with candidate exomoons. 
        Each data point corresponds to one system, with error bars indicating uncertainties in the stellar parameters. 
        The point numbering refers to the stellar identifiers listed in the legend. 
        The Sun is shown for reference (red symbol). 
        Stars with lower effective temperatures ($T_{\rm eff} \lesssim 5800$ K) cover a wide range of ages, 
        while hotter stars are systematically younger, reflecting stellar evolutionary timescales. 
        This distribution provides context for assessing the likelihood of long-term dynamical stability 
        in potential exomoon-hosting systems.
    }
    \label{fig:edad_T}
\end{figure}

Most of the stars with exomoon candidates present solar-like effective temperatures ($\sim$5700 K) and ages between 4–6 Gyr. This agrees with the idea that solar-type stars may provide more favourable conditions for exomoon detection due to observational constraints. For example, techniques such as TTV and TDV are more effective around relatively quiet stars; but on the other hand, the contrast produced in the planet transit around hotter stars is almost imperceptible.
Regarding age, older systems typically host dynamically stable moons, while younger systems may still suffer from dynamical processes that disrupt moon formation or long-term stability.

The similarities in temperature and age between these stars and the Sun may make this sample particularly suitable for exomoon searches from a formation perspective. Nevertheless, this trend could also reflect an observational bias associated with the preference for Sun-like systems in habitability studies.

\subsection{Mercury and Venus: The absence of moons}

The absence of natural satellites around Mercury and Venus stands out as a peculiarity in the otherwise moon-rich Solar System. This contrast raises important questions about the role of dynamical histories and environmental conditions in satellite retention or loss. Although there are several hypotheses for why these two planets have no moons, none of them have been confirmed.

Mercury, the innermost planet, is extremely close to the Sun and has a relatively small mass. These two facts limit its ability to sustain any companion, as most moon configurations would either spiral inwards and crash into the planet due to tidal interactions or be perturbed by the Sun’s gravity and escape from the orbit. These gravitational interactions, preventing the planet from retaining long-term stable moons, are expressed through its Hill sphere, which is significantly smaller than those of the other planets \citep{smart1977textbook}.

On the contrary, Venus is a retrograde and slow rotator, which suggests it had a turbulent dynamical past, possibly resulting from giant impacts or long-term tidal interactions. This uncommon rotational pattern may lead to the instability of a moon orbit, as shown by orbital mechanics simulations \citep{Sheppard_2009}.

Some of the hypotheses proposed to explain the lack of moons around these planets include gravitational disruption by the Sun, since these two planets are the closest to our star. As a result, strong tidal forces may strip away satellites or prevent material from coalescing into stable orbits. Additionally, a hostile formation environment must be considered. Many moons are expected to form through co-formation within circumplanetary disks. However, the inner regions of protoplanetary disks around close-in planets were likely extremely hot and exposed to intense stellar radiation and winds, conditions that may have prevented the retention of the circumplanetary material required for satellite formation.

To assess the possibility of stable moons around Mercury and Venus, we applied the stability criteria detailed in Sect.~\ref{sec:stability_criteria}, namely the Hill radius and the Roche limit. To do this, we assumed a hypothetical satellite with the same mass ratio and orbital distance ratio relative to the planet as the Earth–Moon system. The physical and orbital parameters used for this analysis are listed in Table~\ref{tab:venus_mercury_moons}.

\begin{table}
\caption{Physical and orbital parameters of Mercury and Venus, together with those of a hypothetical moon scaled to Earth–Moon-like mass and semi-major axis ratios. }

\label{tab:venus_mercury_moons}
\centering
\renewcommand{\arraystretch}{1.1}
\small
\begin{tabular}{l@{\hskip 2mm}c@{\hskip 2mm}c@{\hskip 2mm}c@{\hskip 2mm}c@{\hskip 2mm}c@{\hskip 2mm}c}
\hline\hline
Planet & $M_p$ ($M_J$) & $M_m$ ($M_J$) & $R_p$ ($R_J$) & $e_p$ & $a_p$ (AU) & $a_m$ (AU) \\
\hline
Mercury & 0.00017 & 0.000002 & 0.035 & 0.2056 & 0.39 & 0.00099 \\
Venus   & 0.00266 & 0.000033 & 0.087 & 0.0070 & 0.72 & 0.00186 \\
\hline
\end{tabular}
\tablefoot{Planetary masses ($M_p$) and moon masses ($M_m$) are expressed in Jupiter masses ($M_J$), while radii ($R_p$) are expressed in Jupiter radii ($R_J$). Planetary eccentricity ($e_p$) is dimensionless; planetary semi-major axes ($a_p$) are given in AU; and moon semi-major axes ($a_m$) are expressed in AU relative to the host planet.}
\end{table}

Our results, shown in Fig. \ref{fig:SS stb}, show that a hypothetical moon around Venus would satisfy both the Hill and Roche stability conditions, permitting a stable orbit across a wide range of eccentricities ($0 < e_m < 1$). In contrast, a moon around Mercury would violate the Hill criterion for all values of eccentricity, suggesting that stable satellites cannot exist in its environment. This may help explain Mercury's lack of natural satellites. The absence of moons around Venus therefore likely reflects its formation and subsequent dynamical evolution, possibly involving tidal evolution, orbital instabilities, or the loss of primordial satellites during major collisional events.

We additionally verified that both hypothetical satellites satisfy the Laplace criterion by several orders of magnitude. The corresponding Laplace limits ($\sim$36 AU for Mercury and $\sim$26 AU for Venus) are vastly higher than the adopted satellite semi-major axes, indicating that secular stellar perturbations are not the limiting factor in these configurations. Consequently, the Hill criterion remains the dominant dynamical constraint.

\section{Conclusions}
\label{sec:conclusion}

\begin{enumerate}
\item We compiled and homogenised a catalogue of exomoon candidates, including stellar, planetary, and lunar parameters. Several candidates lie near the brown dwarf boundary or may be better explained as ringed planets.
\item We reviewed the main theoretical formation and evolution mechanisms for moons and summarised the current detection techniques (e.g. TTV and TDV), highlighting the strong observational biases that currently restrict detections to large moons in wide orbits.
\item Only 13 candidates in our sample currently have sufficient observational constraints to evaluate dynamical stability in detail. Of these, eight satisfy Hill, Roche, and Laplace stability criteria for low eccentricities, with planetary eccentricity emerging as one of the main limiting factors for long-term stable moon orbits.
\item Dynamical processes such as planetary migration, evection resonance, and close encounters can destabilise moons, providing a plausible explanation for the absence of moons around Mercury and Venus. Recent numerical simulations further show that close scattering encounters can substantially modify moon eccentricities and remove a significant fraction of satellites \citep{Roccetti_2023}.
\item The mass ratios $M_m/M_p$ of most candidates are higher than those in the Solar System, suggesting alternative formation pathways, such as captures or giant impacts.
\item Most host stars have solar-like effective temperatures and ages, consistent with observational biases favouring relatively quiet and photometrically stable stars.
\item By combining dynamical and observational constraints (stellar temperature, age, planetary radius, orbital period, transit geometry, and sky position), we identified a subset of promising candidates for follow-up observations with current and near-future facilities. In particular, 12 dynamically stable candidates fall within the current PLATO long-observation fields (LOPS2 and LOPN1), making them especially relevant targets for future photometric searches.
\item The non-detection of exomoons to date is likely due to a combination of observational limitations, dynamical instability, signal contamination from stellar or planetary variability, and the rarity of favourable formation and evolution conditions. 
Recent re-analyses of previously reported exomoon candidates suggest that many signals may be attributable to stellar variability or instrumental systematics, reinforcing the difficulty of confirming exomoons with current data. 
However, upcoming facilities and techniques --- including long-baseline photometric missions, high-resolution spectroscopy of directly imaged companions, astrometric methods, and high-precision interferometry --- may significantly improve the detectability of large exomoons in the coming years.
Our methodology provides a systematic framework to prioritise targets for future searches, complementing recent efforts focused on future mission detectability and observational strategies for exomoons \citep{limbach_2024_exomoons}. 
Recent studies also suggest that future direct-imaging facilities such as HWO may detect Earth-sized exomoons through mutual-event photometry under favourable observing conditions \citep{limbach_2026_hwo}.
\end{enumerate}

\section*{Data availability}

The full catalogue is available in electronic form in a \href{https://github.com/Aliper02/Exomoons_Catalogue/}{GitHub repository} in order to have it accessible and collaborative. 
The catalogue and visibility analysis presented here provide a practical reference for selecting exomoon targets accessible to future missions such as PLATO, CHEOPS, JWST, and next-generation astrometric facilities.

\begin{acknowledgements}
The authors thank the anonymous referee for their valuable comments that helped improve this manuscript significantly. This research has made use of data obtained from or tools provided by the portal exoplanet.eu of The Extrasolar Planets Encyclopaedia. This research has made use of the SIMBAD database, operated at CDS,
Strasbourg, France. This work benefited from the online platforms \texttt{github} (\url{https://github.com/}) and \texttt{overleaf} . Additionally, this work used the following \textsc{python} packages: \textsc{matplotlib} \citep{Hunter2007}, \textsc{numpy} \citep{Harris2020}, \textsc{pandas} \citep{McKinney2010} and \textsc{astropy} \citep{Astropy2013, Astropy2018, Astropy2022}.
I.R. is supported by the "Atracción de Talento César Nombela" 2025-T1/TEC-36215.
Part of this work was presented as the Master Thesis of A.P.R. in the Universidad Complutense de Madrid in 2025.
\end{acknowledgements}

\bibliographystyle{aa}
\bibliography{bibliography_exomoon}

\onecolumn
\clearpage
\begin{appendix} 

\section{Solar System moon sample}
\label{app:ssmoons}
Table~\ref{tab:solarsystemmoons} compiles the orbital and physical properties of selected Solar System moons used throughout this analysis.

\begin{table} [!t]
\caption{Orbital and physical parameters of 13 Solar System moons considered in the stability analysis. }

\label{tab:solarsystemmoons}
\centering
\begin{tabular}{lccccccccc}
\hline\hline
Moon & Host planet & $M_m$ ($M_J$) & $a_m$ (AU) & $M_p$ ($M_J$) & $e_m$ & $e_p$ & $R_p$ ($R_J$) & $a_p$ (AU) \\
\hline
Moon     & Earth   & $3.9\times10^{-5}$ & 0.002570 & 0.00315 & 0.0549   & 0.0167 & 0.089 & 1.00 \\
Phobos   & Mars    & $5.7\times10^{-12}$ & 0.000063 & 0.00034 & 0.0151   & 0.0934 & 0.048 & 1.52 \\
Deimos   & Mars    & $1.1\times10^{-12}$ & 0.000157 & 0.00034 & 0.0003  & 0.0934 & 0.048 & 1.52 \\
Io       & Jupiter & $4.7\times10^{-5}$  & 0.002820 & 1.00000 & 0.0041   & 0.0480 & 1.000 & 5.20 \\
Europa   & Jupiter & $2.5\times10^{-5}$  & 0.004486 & 1.00000 & 0.0090   & 0.0480 & 1.000 & 5.20 \\
Ganymede & Jupiter & $7.8\times10^{-5}$  & 0.007155 & 1.00000 & 0.0013   & 0.0480 & 1.000 & 5.20 \\
Callisto & Jupiter & $5.7\times10^{-5}$  & 0.012585 & 1.00000 & 0.0074   & 0.0480 & 1.000 & 5.20 \\
Titan    & Saturn  & $7.1\times10^{-5}$  & 0.008168 & 0.29942 & 0.0288   & 0.0560 & 0.833 & 9.58 \\
Rhea     & Saturn  & $1.2\times10^{-6}$  & 0.003523 & 0.29942 & 0.00125 & 0.0560 & 0.833 & 9.58 \\
Titania  & Uranus  & $1.9\times10^{-6}$  & 0.002916 & 0.04574 & 0.0022   & 0.0470 & 0.363 & 19.23 \\
Oberon   & Uranus  & $1.6\times10^{-6}$  & 0.003900 & 0.04574 & 0.0008   & 0.0470 & 0.363 & 19.23 \\
Triton   & Neptune & $1.1\times10^{-5}$  & 0.002371 & 0.05395 & 0.00001 & 0.0113 & 0.352 & 30.10 \\
Proteus  & Neptune & $2.6\times10^{-8}$  & 0.000786 & 0.05395 & 0.0005   & 0.0113 & 0.352 & 30.10 \\
\hline
\end{tabular}
\tablefoot{This table lists the host planet, the moon mass $M_m$ (in Jupiter masses, $M_J$), the semi-major axis of the moon’s orbit $a_m$ (in astronomical units, AU), the planetary mass $M_p$ (in $M_J$), the eccentricities of the moon ($e_m$) and host planet ($e_p$), the planetary radius $R_p$ (in Jupiter radii, $R_J$), and the planet’s orbital semi-major axis $a_p$ (AU).}
\tablebib{\citet{oan2026anuario}.}
\end{table}

\section{Comparison with Solar System moons}

To characterise the moon population considered in this work, we plotted together the Solar System moon sample and the exomoons listed in Table \ref{tab:full_catalogue} based on the mass ratio with the host planet and the eccentricity of their orbits. The exomoon values should be interpreted cautiously, as many are inferred from indirect observational constraints and remain highly uncertain.

\begin{figure} [!t]
    \centering
    \includegraphics[width=\textwidth]{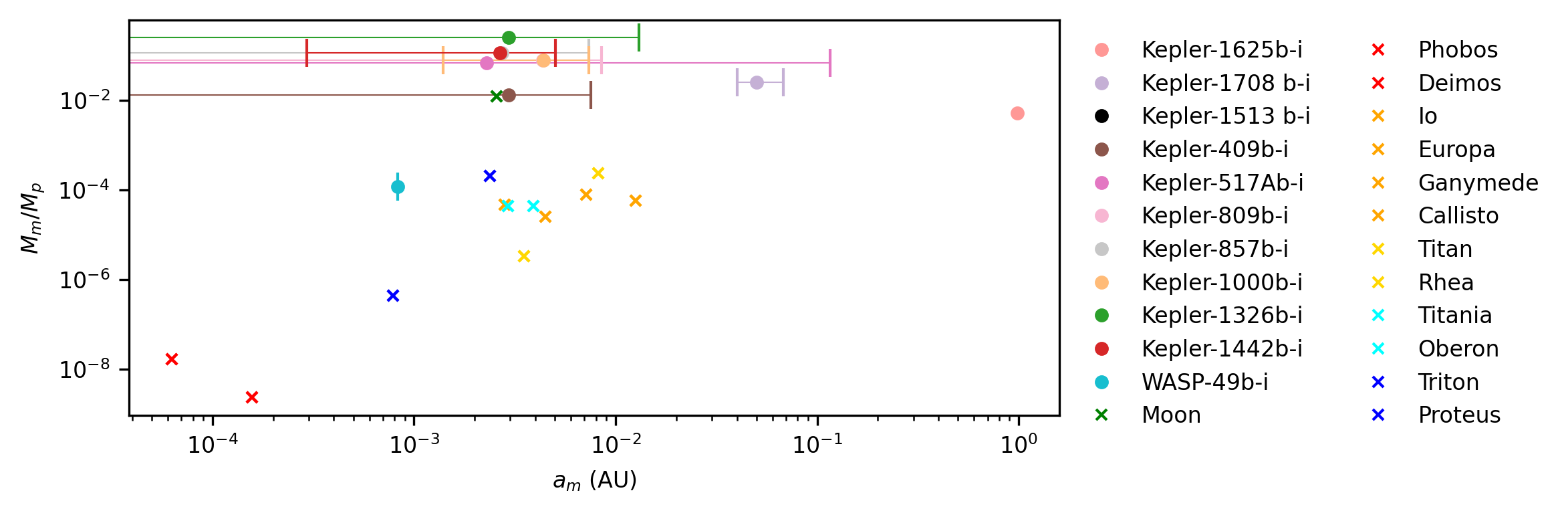} 
    \caption{Moon-to-planet mass ratio ($M_m/M_p$) as a function of the moon’s semi-major axis ($a_m$) for our Solar System moon sample (crosses) and exomoon candidates (dots). Each Solar System planet’s satellites are plotted in the same colour for clarity. Error bars are included for exomoon candidates to account for observational or theoretical uncertainties. The distribution shows that Solar System moons populate a wide range of semi-major axes with relatively small mass ratios, while the exomoon candidates cluster at significantly larger values of $M_m/M_p$, reinforcing the detection bias observed in Fig.~\ref{fig:histo}. This comparison underlines the contrast between the parameter space accessible to current exomoon searches and the actual population of moons in our Solar System.}
    \label{fig:exomoons ratio}  
\end{figure}

\begin{figure} [!t]
    \centering
    \includegraphics[width=\textwidth]{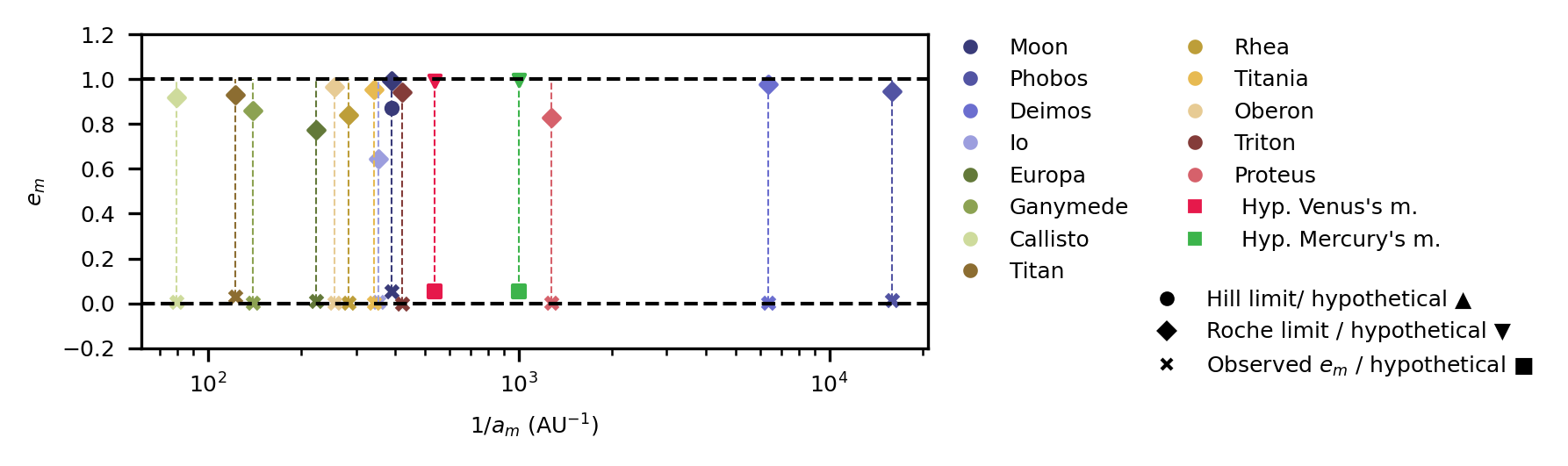} 
 \caption{
Orbital eccentricities of major moons in the Solar System (crosses) compared against analytical stability criteria. 
The dots represent the eccentricity upper limits derived from the Hill criterion (\( e_m < R_H/a_M - 1\)) and the diamonds, from the Roche criterion (\( e_m < -R_H/a_M + 1\)), calculated individually for each system. 
All Solar System moons lie comfortably below these critical thresholds, confirming that none is at risk of tidal disruption or dynamical ejection under present-day conditions. Hill stability limits not explicitly plotted exceed unity, implying that any eccentricity value would be dynamically permissible. For Mercury and Venus, the hypothetical moons (squares) have Hill limits (triangles) of 2.3915 (Venus) and -0.4293 (Mercury). Points that do not appear in the plot correspond to stability limits above 1 (dynamically allowed) or below 0 (dynamically forbidden).} 
    \label{fig:SS stb}  
\end{figure}

\end{appendix}

\end{document}